\documentclass[a4paper,twoside,prl,showpacs,preprintnumbers,twocolumn,superscriptaddress,nofootinbib]{revtex4-2}

\usepackage{amssymb,amsmath,bm,natbib}
\usepackage{color}
\usepackage{slashed}
\usepackage{graphics}
\usepackage{graphicx}
\usepackage[utf8]{inputenc}
\usepackage[caption=false]{subfig}
\usepackage{hyperref}
\usepackage{url}
\usepackage{dsfont}
\usepackage{float} 
\usepackage{cancel}
\usepackage{xcolor,colortbl}
\usepackage{rotating}
\usepackage[export]{adjustbox}
\usepackage{subfig}
\usepackage{cleveref}
\usepackage{booktabs}
\usepackage{multirow}
\usepackage{braket}
\usepackage{float}
\usepackage[shortlabels]{enumitem}
\newcommand{\GeV}{{\rm\ GeV}}
\newcommand{\TeV}{{\rm\ TeV}}
\newcommand{\fb}{{\rm\ fb}}
\newcommand{\Lag}{{\mathcal L}}

\newcommand{\SUD}{{SU(2)_{\rm D}}}
\newcommand{\Z}{{\mathbb Z}}

\begin{document}

\title{A fermionic portal to a non-abelian dark sector}

\author{Alexander Belyaev}
\email{a.belyaev@soton.ac.uk}
\affiliation{School of Physics and Astronomy, University of Southampton, Highfield, Southampton SO17 1BJ, UK}

\author{Aldo Deandrea}
\email{deandrea@ipnl.in2p3.fr}
\affiliation{Univ. Lyon, Universit{\' e} Claude Bernard Lyon 1, CNRS/IN2P3,
 IP2I UMR5822, F-69622, Villeurbanne, France}

\author{Stefano Moretti}
\email{s.moretti@soton.ac.uk; stefano.moretti@physics.uu.se}
\affiliation{School of Physics and Astronomy, University of Southampton, Highfield, Southampton SO17 1BJ, UK}
\affiliation{Department of Physics and Astronomy, Uppsala University, Box 516, SE-751 20 Uppsala, Sweden}

\author{Luca Panizzi}
\email{luca.panizzi@physics.uu.se}
\affiliation{School of Physics and Astronomy, University of Southampton, Highfield, Southampton SO17 1BJ, UK}
\affiliation{Department of Physics and Astronomy, Uppsala University, Box 516, SE-751 20 Uppsala, Sweden}

\author{Nakorn Thongyoi}
\email{nakorn.thongyoi@gmail.com}
\affiliation{School of Physics and Astronomy, University of Southampton, Highfield, Southampton SO17 1BJ, UK}

\begin{abstract}
We introduce a new class of renormalisable models, consisting of a dark $\SUD$ gauge sector connected to the Standard Model (SM) through a Vector-Like (VL) fermion mediator, not requiring a Higgs portal, in which a massive vector boson is the Dark Matter (DM) candidate.
These models are labelled Fermion Portal Vector Dark Matter (FPVDM).
Multiple realisations are possible, depending on the properties of the VL partner and of the scalar potential. One example is discussed in detail. 
FPVDM models have a large number of applications in collider and non-collider experiments, depending on the mediator sector. 
\end{abstract}

\maketitle


The nature of DM, which 
existence has been established beyond any reasonable doubt  by several independent cosmological observations,
is one of the greatest puzzles of contemporary particle physics.
Models with a vector DM, especially in the non-abelian case,
are the least explored but well motivated, as the gauge principle offers guidance  and constraints limiting the possible theoretical constructions (see, e.g.,  \cite{Hambye:2008bq,Chen:2009ab,DiazCruz:2010dc,Bhattacharya:2011tr,Koorambas:2013una,Fraser:2014yga,Hubisz:2004ft,Huang:2015wts,Ko:2016fcd,Barman:2017yzr,Huang:2017bto,Barman:2018esi,Barman:2019lvm,Abe:2020mph,Chowdhury:2021tnm,Baouche:2021wwa} 
for a discussion of non-abelian DM  in different set-ups, in particular using non-renormalisable kinetic mixing terms or Higgs portal scenarios). 
In this letter we suggest a new framework which extends the gauge sector of the Standard Model (SM) 
by a new non-abelian gauge group for which no renormalisable kinetic mixing terms are allowed\footnote{Contributions to gauge kinetic mixing may arise at loop level, depending on the structure of the Higgs sector, but they correspond to suppressed higher operator terms.} and under which all SM particle are singlets. The simplest non-abelian group is $SU(2)$, which in the following will be labelled $\SUD$ as it connects the SM to the dark sector. The gauge bosons associated to $\SUD$ are labelled as $V_\mu^D=\left(V^0_{D+\mu}~V^0_{D0\mu}~V^0_{D-\mu}\right)$, where, here and in the following, the electric charge is specified in the fields superscripts, while the isospin under $\SUD$ (D-isospin) is specified in the fields subscripts. 
The covariant derivative associated with $\SUD$ is:
\begin{eqnarray}
\label{eq:covdev}
 D_\mu &=& \partial_\mu - \left( i {g_{D} \over \sqrt 2} V^0_{{D\pm}\mu} T^\pm_{D} + i g_{D} V^0_{{D0}\mu} T_{{3D}} \right)\;,
\end{eqnarray}
where $g_{D}$ is the $\SUD$ coupling constant and $T_{{3D}}$ is the D-isospin. 

The fields responsible for  breaking  the gauge symmetries are two scalar doublets:
\begin{eqnarray}
\begin{array}{cclcccl}
 \Phi_H   &=& \left(\phi^+ \; \phi^0 \right)^T &\rightsquigarrow& 
\langle \Phi_H \rangle &=& {1\over\sqrt{2}} \left(0 \; v \right)^T, \label{eq:Higgsdoublet} \\
 \Phi_{D} &=& \left(\varphi^0_{D+{1\over2}} \; \varphi^0_{D-{1\over2}} \right)^T &\rightsquigarrow&
\langle \Phi_{D} \rangle &=& {1\over\sqrt{2}} \left(0 \; v_{D} \right)^T,\label{eq:HDdoublet}
\end{array}
\end{eqnarray}
where the first is breaking $SU(2)_L\times U(1)_Y$, while the second is breaking $\SUD$ via their respective Vacuum Expectation Values (VEVs) $v$ and $v_{D}$. 

A $\Z_2$ symmetry is then introduced to stabilise the DM candidate associated with the lightest $\Z_2$-odd particle of the model. It is assumed that different members of the same $\SUD$ multiplet have a different $\Z_2$ parity, while all SM states are even under $\Z_2$. The fact that SM fields are neutral under the D-charge has a remarkable consequence: the
lightest D-charged and electrically-neutral state is stable and is therefore a dark matter
candidate. 

The possible origin of the $\Z_2$ parity can be, for example, linked to an additional $U(1)_{D}$ which would also introduce a kinetic mixing between $U(1)_Y$ and $U(1)_{D}$. 
Another possibility is the underlying strongly-coupled sector which condensates form particles in the low energy regime.
A detailed discussion of this case is given in \cite{Ma:2015gra} and further used in \cite{Wu:2017iji} for a scalar DM candidate. 

A DM candidate which is charged under a new dark gauge group can potentially induce a phase of dark matter genesis in the early universe, as discussed in the $U(1)_D$ case in~\cite{Jimenez:2020bgw}, which can be used to explain the tension in the measurement of the Hubble constant from Planck and supernovae data. We do not explore this interesting possibility as it requires a dedicated study and goes beyond the minimal model set-up we wish to discuss here. 

The connection between the SM and the dark sector is provided by two new VL fermions, which are singlets of $SU(2)_L$ but form a doublet under $\SUD$,
$\Psi=(\psi_{D} \;\psi )$.\footnote{VL portals have also been explored in \cite{Baek:2017ykw,Colucci:2018vxz}, but for scalar DM candidates.} This fermion doublet contains a $\Z_2$-odd component and a $\Z_2$-even component, which without loss of generality can be identified with the $T_{{3D}}=+1/2$ and $T_{{3D}}=-1/2$ D-isospin components, respectively:  the latter can mix with SM fermions which share the same SM quantum numbers. The mass and interaction Lagrangian of the fermion sector is:
\begin{equation}
 -\Lag_f = M_\Psi \bar \Psi \Psi + (y \bar f^{\rm SM}_L \Phi_H f^{\rm SM}_R + y^\prime \bar \Psi_L \Phi_{D} f^{\rm SM}_R + h.c)\;,
\end{equation}
where $f^{\rm SM}_{L,R}$ generically denotes a SM left-handed doublet or right-handed singlet, $y$ is the Yukawa coupling of the SM and $y^\prime$ is a new Yukawa coupling connecting the SM fermion with $\Psi$. The particle content of the model, which we abbreviate as FPVDM,  is summarised in tab. \ref{tab:particlesQN}.

{
\setlength{\tabcolsep}{3pt}
\setlength{\arraycolsep}{0pt}
\begin{table}[h]
\centering
\begin{tabular}{c|cc|c||c}
\hline
&&&&\\[-8pt]
 & $SU(2)_L$ & $U(1)_Y$ & $\SUD$ & $\Z_2$ \\
&&&&\\[-9pt]
\hline
&&&&\\[-8pt]
$\Phi_{D}=\left(\begin{array}{c} \varphi^0_{D+{1\over2}} \\ \varphi^0_{D-{1\over2}} \end{array}\right)$ & $\mathbf{1}$ & $0$ & $\mathbf{2}$ & $\begin{array}{c} - \\ + \end{array}$ \\[10pt]
\hline
&&&&\\[-8pt]
\multirow{2}{*}{$\Psi=\left(\begin{array}{c} \psi_{D} \\ \psi \end{array}\right)$} & \multirow{2}{*}{$\mathbf{1}$} & \multirow{2}{*}{$Q$} & \multirow{2}{*}{$\mathbf{2}$} & $-$ \\
& & & & $+$\\[2pt]
\hline
&&&&\\[-8pt]
$V_{D\mu}=\left(\begin{array}{c} V^0_{D+\mu} \\ V^0_{D0\mu} \\ V^0_{D-\mu} \end{array}\right)$ & $\mathbf{1}$ & $0$ & $\mathbf{3}$ & $\begin{array}{c} - \\ + \\ - \end{array}$ \\[12pt]
\hline
\end{tabular}
\caption{\label{tab:particlesQN}The quantum numbers  of the new particles under the Electro-Weak (EW) and $\SUD$ gauge groups.}
\end{table}
}

\noindent The scalar potential for $\Phi_H$ and $\Phi_{D}$ reads:
\begin{eqnarray}
 V(\Phi_H,\Phi_{D}) &=& - \mu^2 \Phi_H^\dagger \Phi_H + \lambda (\Phi_H^\dagger \Phi_H)^2 - \mu_{D}^2 \Phi_{D}^\dagger \Phi_{D} \nonumber\\
 &+& \lambda_{D} (\Phi_{D}^\dagger \Phi_{D})^2 + \lambda_{\Phi_H\Phi_{D}} \Phi_H^\dagger \Phi_H \; \Phi_{D}^\dagger \Phi_{D}\;.
 \label{eq:scalarpotential}
\end{eqnarray} 
This potential has non-trivial stationary points at 
\begin{equation}
\left\{\begin{array}{l}
v=\pm\sqrt{4\lambda_{D} \mu^2-2\lambda_{\Phi_H\Phi_{D}}\mu_{D}^2 \over 4\lambda\lambda_{D}-\lambda_{\Phi_H\Phi_{D}}^2} \\
v_{D}=\pm\sqrt{4\lambda \mu_{D}^2 - 2\lambda_{\Phi_H\Phi_{D}}\mu^2 \over 4\lambda\lambda_{D}-\lambda_{\Phi_H\Phi_{D}}^2}
\end{array}\right.\;, 
\end{equation}
which define its minima if the following conditions are satisfied: $\mu \neq 0$, $\mu_{D} \neq 0$ and either $\{\lambda_{\Phi_H\Phi_{D}}<0$, $\lambda>0$, $\lambda_{D}>0$,  $\lambda_{\Phi_H\Phi_{D}}^2<4 \lambda \lambda_{D}\}$ or $\{\lambda_{\Phi_H\Phi_{D}}>0$, $2 \lambda \mu_{D}^2 > \lambda_{\Phi_H\Phi_{D}} \mu^2$,  $2 \lambda_{D} \mu^2 > \lambda_{\Phi_H\Phi_{D}} \mu_{D}^2\}$.


The theory contains 6 massive gauge bosons ($Z$, $W^\pm$, $V^0_{D0}$ and $V^0_{D\pm}$) and therefore 6 Goldstone bosons correspond to their longitudinal components. The remaining 2 degrees of freedom correspond to physical scalars, which include the Higgs boson of the SM and another CP-even scalar. By denoting the neutral scalars in terms of their components in the unitary gauge as $
\phi^0 = {1\over\sqrt{2}} (v + h_1)$ and $\varphi^0_{D-1/2} = {1\over\sqrt{2}} (v_{D} + \varphi_1)$, the mass terms of the scalar Lagrangian reads:
\begin{equation}
\Lag_m^{\mathcal S} = 
(h_1 \; \varphi_1) 
\left(
\begin{array}{cc}
\lambda v^2 & {\lambda_{\Phi_H\Phi_{D}}\over2} v v_{D} \\ {\lambda_{\Phi_H\Phi_{D}}\over2} v v_{D} & \lambda_{D} v_{D}^2
\end{array}
\right)
\left( \begin{array}{c} h_1 \\ \varphi_1 \end{array} \right) \;.
\end{equation}
The mass eigenvalues can be obtained by diagonalising the mass matrix via a rotation matrix $V_S=\left(\begin{array}{cc} \cos\theta_S & \sin\theta_S \\ -\sin\theta_S & \cos\theta_S \end{array} \right)$ and read:
\begin{equation}
 m_{h,H_{D}}^2=\lambda v^2+\lambda_{D} v_{D}^2\mp\sqrt{(\lambda v^2-\lambda_{D} v_{D}^2)^2+\lambda_{\Phi_H\Phi_{D}}^2v^2 v_{D}^2}\;,
\end{equation}
with mixing angle $\sin\theta_S = \sqrt{2{m_{H_{D}}^2 v^2 \lambda - m_h^2 v_{D}^2 \lambda_{D} \over m_{H_{D}}^4 - m_h^4}}$.

The masses of the SM gauge bosons are not altered by the presence of $\Phi_{D}$. 
The gauge bosons of $\SUD$ are all degenerate in mass at tree level:
\begin{equation}
m_{V_{D}}\equiv m_{V^0_{D\pm}} = m_{V^0_{D0}} = {g_{D}\over2} v_{D} \label{eq:VPmass} \;.
\end{equation}
This degeneracy is broken by the different fermionic loop corrections associated with the opposite $\Z_2$ parities of the $\SUD$ gauge bosons. 
{The only electrically neutral and massive $\Z_2$-odd states of FPVDM scenarios are the $\SUD$ gauge bosons $V^0_{D\pm}$ which are therefore the DM candidates.\footnote{In principle, the introduction of VL neutrino partners could make the $\Z_2$-odd member of the VL doublet a DM candidate, but a mixing in the neutrino sector would require further new physics to account for neutrino mass generation. This non-minimal scenario is not considered in the present analysis.}}

In the fermion sector, the component with $T_{{3D}}=1/2$ gets only a VL mass,  therefore 
\begin{equation}
m_{\psi_{D}}=M_\Psi\;,
\end{equation}
while the other fermion masses are generated after both scalars acquire a VEV. The fermionic mass matrix reads:
{\setlength{\arraycolsep}{0pt}
\begin{equation}
 \Lag_m^f = (\bar f^{\rm SM}_L \psi_L) \mathcal M_F \left(\begin{array}{c} f^{\rm SM}_R \\ \psi_R \end{array}\right),\;\text{with}\;\mathcal M_F = \left(\begin{array}{cc} y {v\over\sqrt 2} & 0 \\ y^\prime {v_{D}\over\sqrt 2} & M_\Psi \end{array}\right)\;.
\end{equation}}
This mass matrix describes the mixing of a VL fermion with a SM fermion driven by $\Phi_{D}$. The mass matrix can be diagonalised by two unitary matrices, $V_{L,R}$, leading to the mass eigenstates $f$ and $F$, where $f$ identifies the SM fermion and $F$ its heavier partner:
{\setlength{\tabcolsep}{3pt}
\setlength{\arraycolsep}{0pt}
\begin{equation}
 \Lag_m^f = (\bar f_L F_L) \mathcal M_F^d \left(\begin{array}{c} f_R \\ F_R \end{array}\right) = (\bar f_L F_L) V_{fL}^\dagger \mathcal M_F V_{fR} \left(\begin{array}{c} f_R \\ F_R \end{array}\right)\;,
\end{equation}}
with $V_{fL,R}=\left(\begin{array}{cc} \cos\theta_{fL,R} & \sin\theta_{fL,R} \\ -\sin\theta_{fL,R} & \cos\theta_{fL,R} \end{array} \right)$. The mass eigenvalues are:
\begin{eqnarray}
 m_{f,F}^2&=&{1\over4} \bigg[y^2 v^2 + y^{\prime2} v_{D}^2 + 2 M_\Psi \nonumber\\
 &\mp& \sqrt{(y^2 v^2 + y^{\prime2} v_{D}^2 + 2 M_\Psi)^2-8y^2v^2M_\Psi^2}\bigg]\;.
\end{eqnarray}
The fermion sector contains therefore a SM fermion with mass $m_f$, a $\Z_2$-even partner with mass $m_F$ and a $\Z_2$-odd partner with mass $m_{\psi_{D}}$. The mass hierarchy is $m_f<m_{\psi_{D}}\leq m_F$.

The Yukawa parameters can be traded for the masses of the physical fermions as 
\begin{equation}
y = \sqrt{2} {m_f m_F\over m_{\psi_{D}} v},~
y^\prime = \sqrt2 {\sqrt{(m_F^2 - m_{\psi_{D}}^2)(m_{\psi_{D}}^2 - m_f^2)}\over m_{\psi_{D}} v_{D}} 
\end{equation}
and the mixing angles as
\begin{equation}
\sin^2\theta_{fL} = {m_f^2\over m_{\psi_{D}}^2} {m_F^2 - m_{\psi_{D}}^2 \over m_F^2 - m_f^2},~
\sin^2\theta_{fR} = {m_F^2 - m_{\psi_{D}}^2 \over m_F^2 - m_f^2}\;.
\end{equation}
The left-handed mixing angle is suppressed by the ${m_f^2/m_{\psi_{D}}^2}$ ratio. 
The new fermion sector is completely decoupled in the limit $m_F=m_{\psi_{D}}$, for which $y=y_{\rm SM}=\sqrt2{m_f\over v}$, $y^\prime=0$, $\sin\theta_{fL}=\sin\theta_{fR}=0$, so that the pure SM scenario is restored.

When the full flavour structure of the SM is taken into consideration, different possibilities can be considered. A VL fermion can interact with one or more SM flavours, plus there can be multiple VL fermions. The Cabibbo-Kobayashi-Maskawa (CKM) matrix of the SM might also receive contributions from new physics induced by the mixing of SM and VL quarks. 
In the following we assume that new VL fermions interact only with one SM flavour: in this case all the Lagrangian parameters can be traded for the masses of the physical states, the EW coupling constant $g$ (or equivalently, the fine structure constant $\alpha_{\rm EM}$), the new gauge coupling $g_{D}$, the mixing angle between the scalar fields $\theta_S$ and the measured CKM parameters. Six independent input parameters are thus necessary to describe the new physics sector of the model, namely:
$g_{D}$, $M_{V_{D}}$, $m_{H_{D}}$, $\sin\theta_S$, $m_F$, $m_{\psi_{D}}$. Notice that if there is no mixing in the scalar sector ($\theta_S=0$), there is no Higgs portal at tree level.

Let us now discuss a specific realisation of the model, assuming only one VL partner interacting exclusively with the SM top quark and no mixing between $h$ and $H_{D}$, {i.e.}, $\theta_S=0$. This choice significantly simplifies the Lagrangian: the Higgs sector of the SM is not affected by the new physics at tree-level and the potential of $\Phi_{D}$ has the very same structure as the Higgs potential. A mixing between $h$ and $H_{D}$ is induced only by fermionic loops and will be neglected in the following. The hierarchy between the masses in the fermion sector is $m_t < m_{t_{D}} \leq m_T$, while $H_{D}$ can have any mass allowed by experimental bounds, including lighter than the SM Higgs boson.
These choices are dictated, on the one hand, by minimality and, on the other hand, by a scenario where a non-abelian dark sector is connected to the SM exclusively through the fermion sector without a Higgs portal.

In our study we test this realisation of the model against multiple observables from cosmology, DM Direct and Indirect Detection (DD and ID) experiments and LHC searches. For this purpose the Lagrangian has been implemented in 
{\sc LanHEP}~\cite{Semenov2009}  and 
{\sc FeynRules}~\cite{Alloul:2013bka} while 
model files have been generated in {\sc CalcHEP}~\cite{Belyaev:2012qa}, {\sc UFO}~\cite{Degrande:2011ua} as well as {\sc FeynArts}~\cite{Hahn:2000kx} formats and are available 
on the {\sc HEPMDB}~\cite{hepmdb}. 
This implementation has been used in {\sc micrOMEGAs v5.2.7}~\cite{Belanger:2020gnr} for the evaluation of various DM observables and for extracting the respective limits.
The model implementation in {\sc UFO} format has been used in {\sc MG5\_aMC}~\cite{Alwall:2014hca} for the determination of the LHC constraints. Collider simulations have been performed at LO using the NNPDF3.0 LO set~\cite{NNPDF:2014otw} through the {\sc LHAPDF6} library~\cite{Buckley:2014ana} (LHA index 262400). A simplified version of the model has been implemented to calculate cross-sections at one loop in {\sc MG5\_aMC}
and {\sc FormCalc9.8}~\cite{Hahn:2016ebn}.

\begin{figure}[h]
\centering
\begin{minipage}{.16\textwidth}
	\hspace*{-22pt}\includegraphics[width=0.7\textwidth]{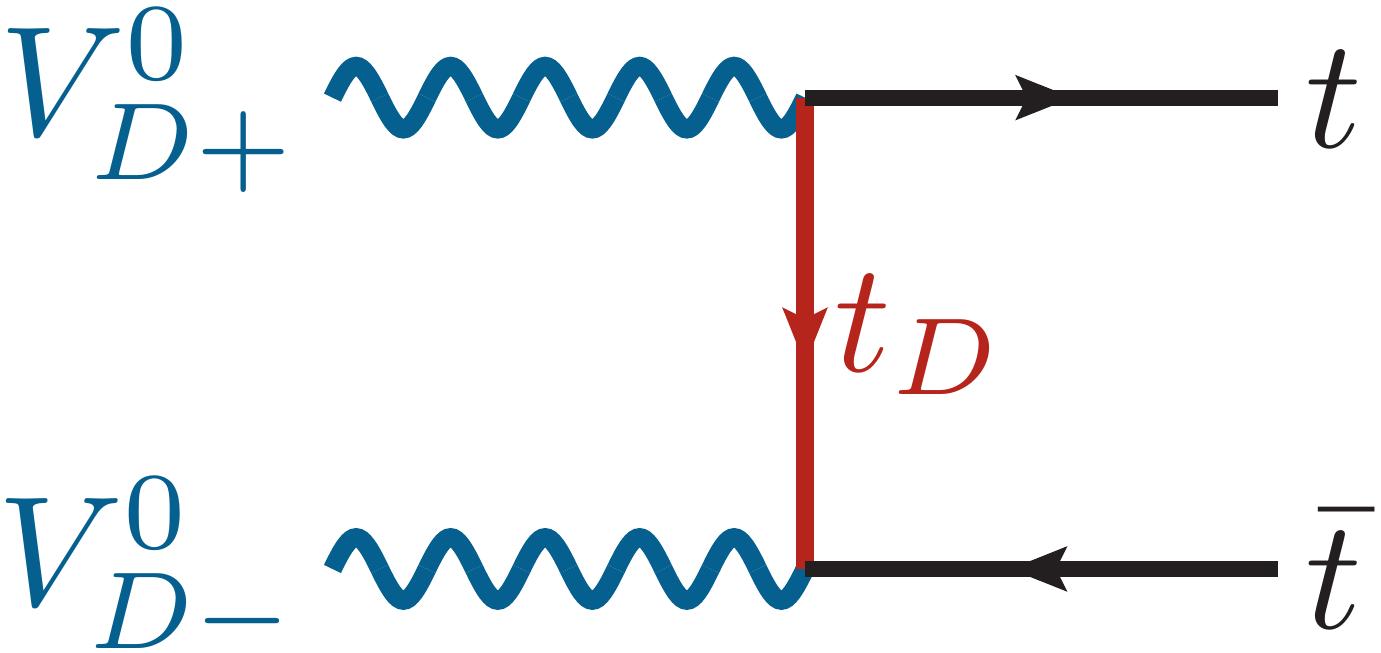}\\[5pt]
	\hspace*{-10pt}\includegraphics[width=0.9\textwidth]{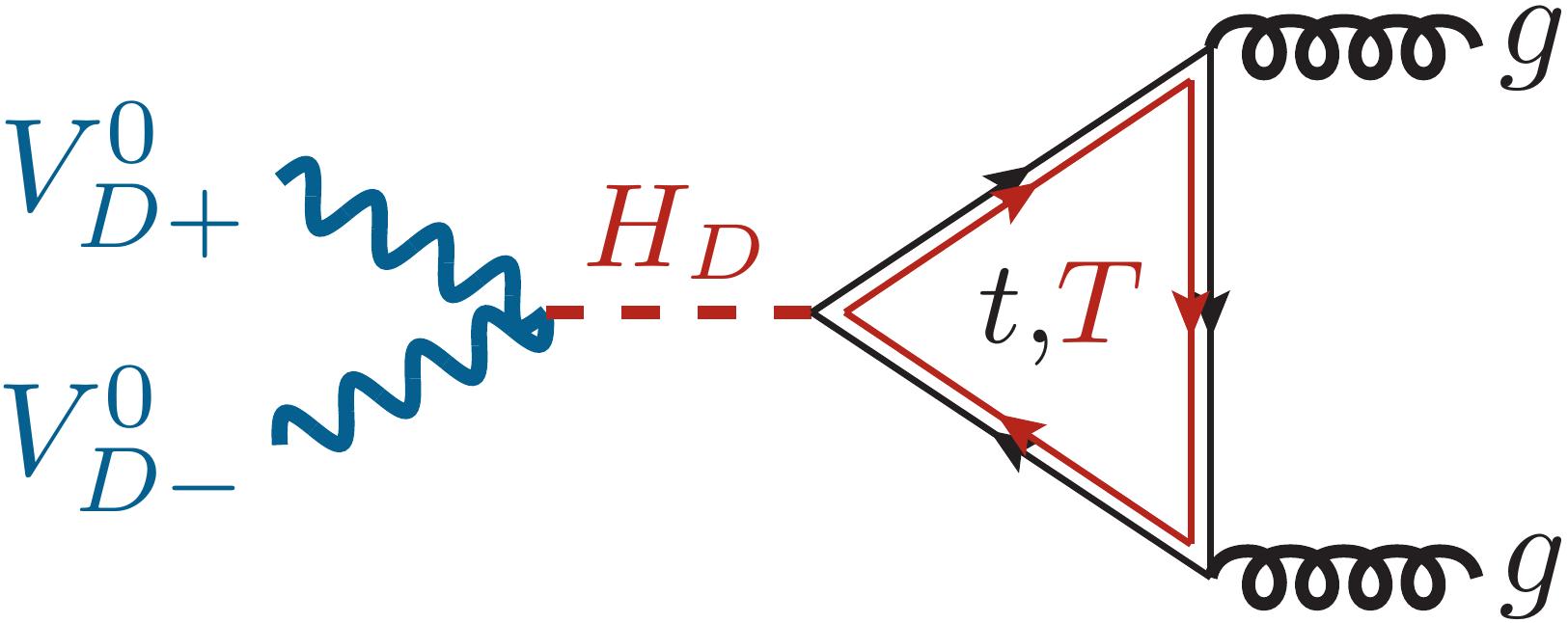}\\[5pt]
	\hspace*{-10pt}\includegraphics[width=0.9\textwidth]{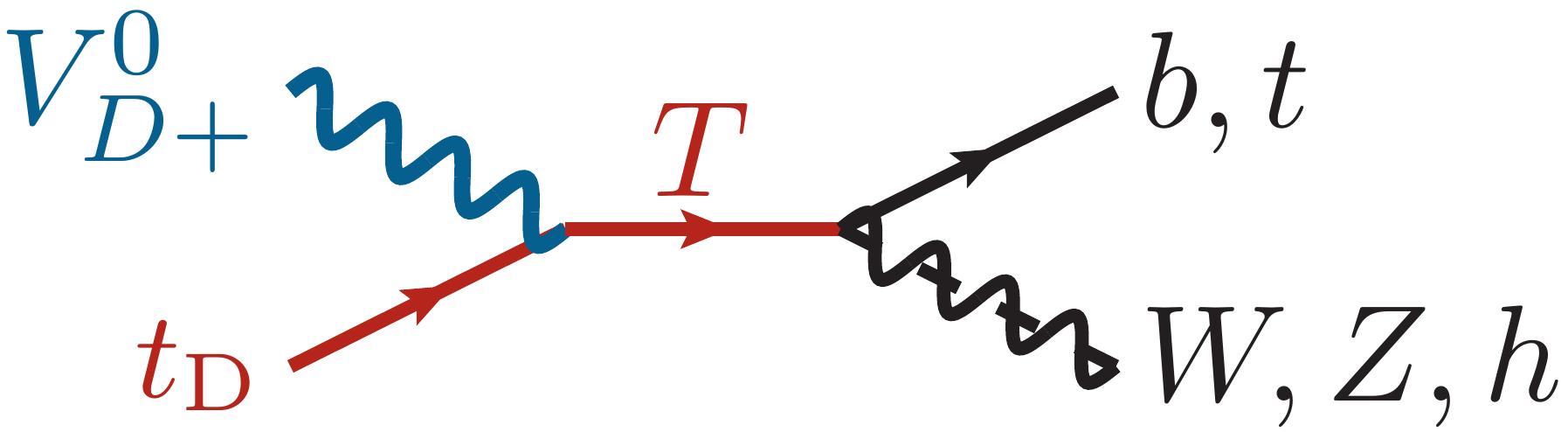}
\end{minipage}
\begin{minipage}{.13\textwidth}
	\hspace*{-5pt}\includegraphics[width=\textwidth]{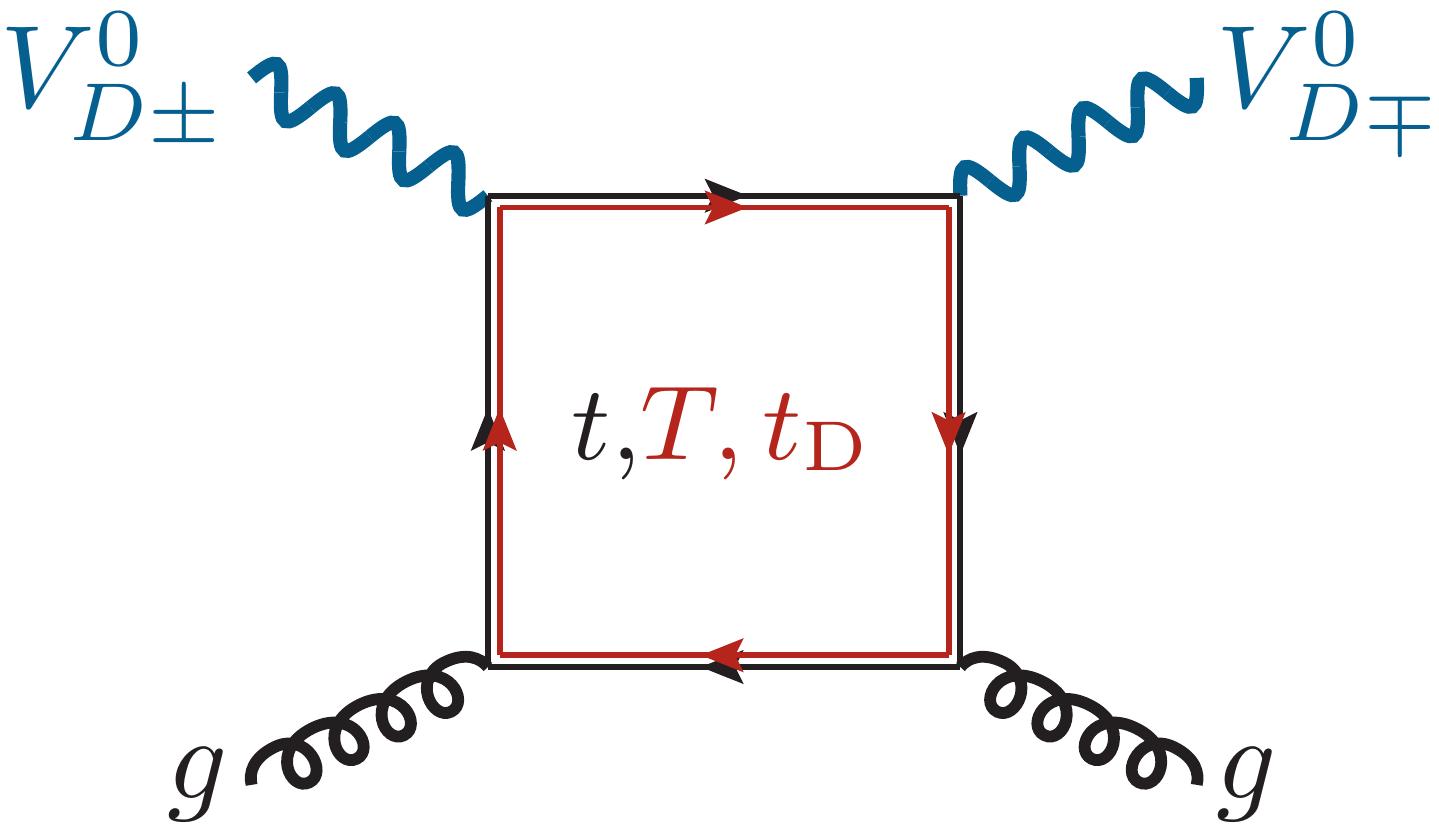}\\[5pt]
	\hspace*{-5pt}\includegraphics[width=0.8\textwidth]{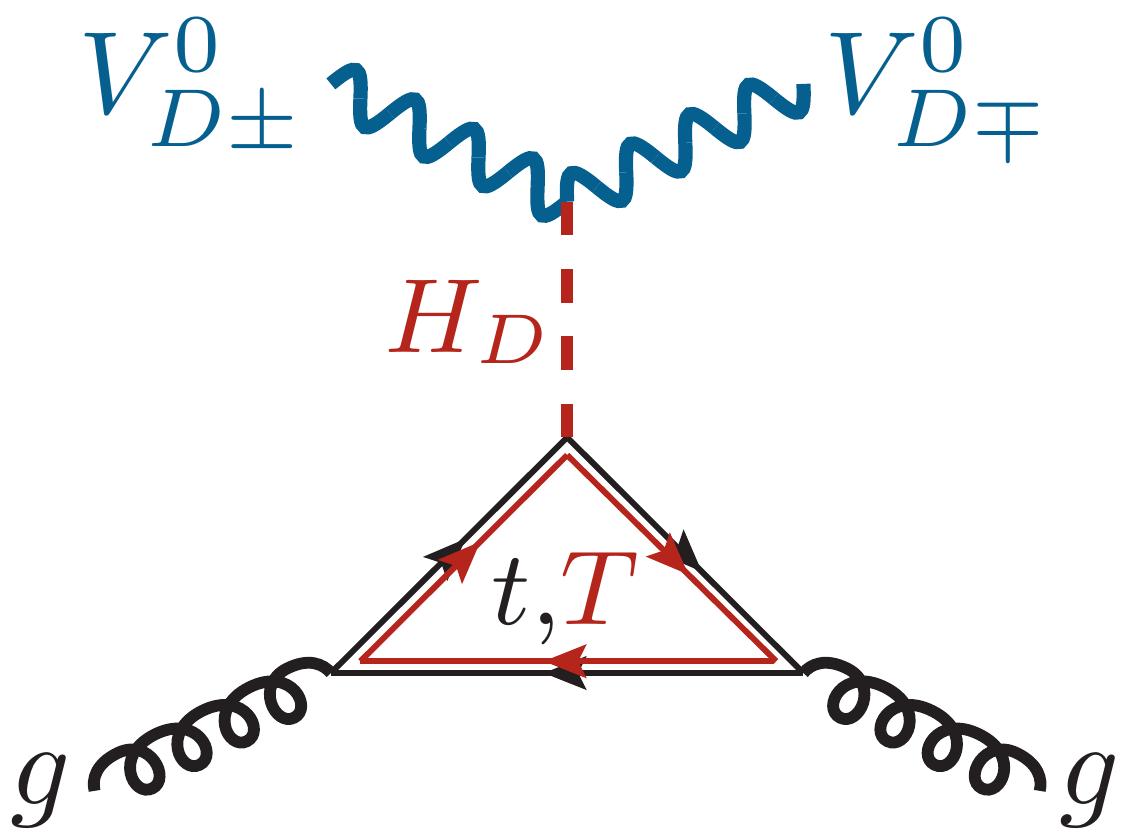}
\end{minipage}
\begin{minipage}{.17\textwidth}
    \hspace*{0pt}\includegraphics[width=\textwidth]{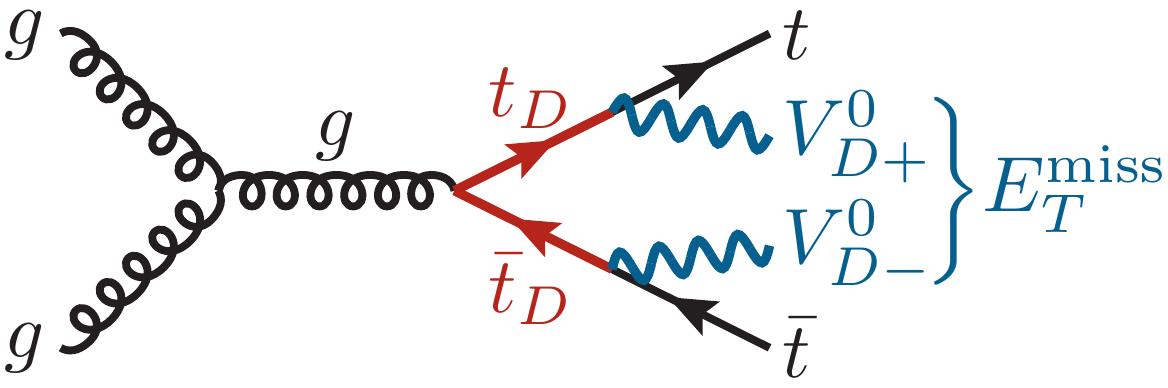}\\[5pt]
    \hspace*{-10pt}\includegraphics[width=.8\textwidth]{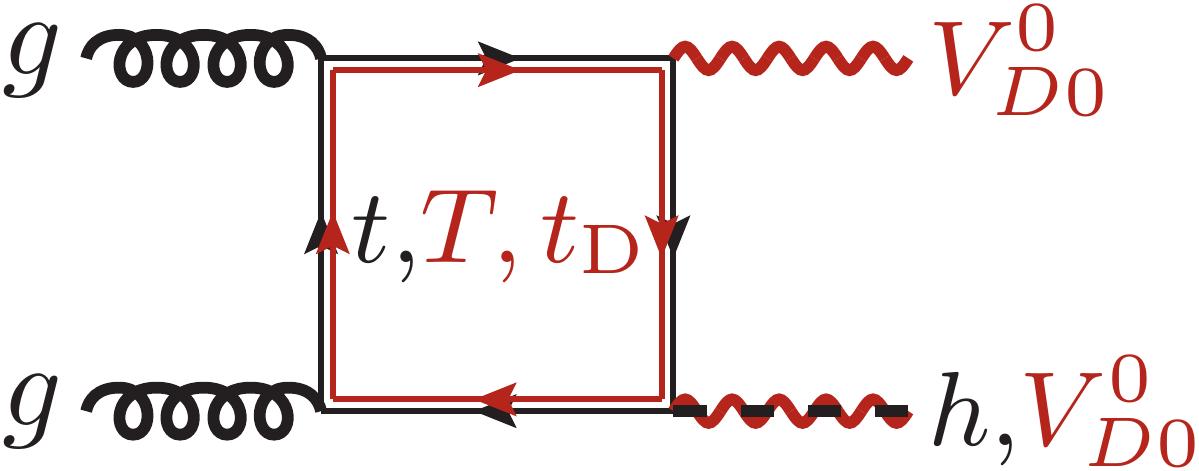}\\
\end{minipage}
\\[3pt]
\begin{minipage}{.13\textwidth}
\hspace*{-18pt}{(a) Relic density and ID}
\end{minipage}
\begin{minipage}{.16\textwidth}
\hspace*{-6pt}{(b) Direct detection}
\end{minipage}
\begin{minipage}{.14\textwidth}
\hspace*{-30pt}{(c) LHC}
\end{minipage}

\caption{\label{fig:feynmandiagrams} (a) $t$-channel and resonant contributions to DM annihilation and DM-mediator co-annihilation processes. (b) Representative diagrams for DD processes. (c) Production processes at the LHC: $t\bar t$ + $E_T^{\rm miss}$, $h V^0_{D0}$ and $V^0_{D0}V^0_{D0}$. }
\end{figure}

The measured amount of relic density is determined by the interplay of annihilation and co-annihilation processes shown in \cref{fig:feynmandiagrams}(a).
ID constraints are associated with DM annihilation rates at CMB time, excluding  regions of parameter space where the injection into SM-plasma in the early Universe is too large
to be consistent with CMB data.
Both relic density and ID processes are tested against PLANCK data~\cite{Planck:2018vyg}.

DD processes are represented by the diagrams in \cref{fig:feynmandiagrams}(b) and tested against limits from XENON 1T~\cite{XENON:2018voc}.

The LHC bound has been obtained via testing of $t_{D}$ pair production with subsequent decay into $V^0_{D\pm}$ and top quarks against CMS searches for top squark pair production decaying into DM. These analyses were done for the final states with opposite sign leptons and missing transverse energy $E_T^{\rm miss}$~\cite{CMS:2017jrd}, recast through the {\sc MadAnalysis 5} framework~\cite{Conte:2018vmg,DVN/BQM0T3_2021}. We also estimated the relevance of $V^0_{D0}$ pair production and associate production of $V^0_{D0}$ with the Higgs boson, occurring at LO via fermion loops. The tested processes are shown in \cref{fig:feynmandiagrams}(c).

\begin{figure}[h]
\includegraphics[width=.5\textwidth]
{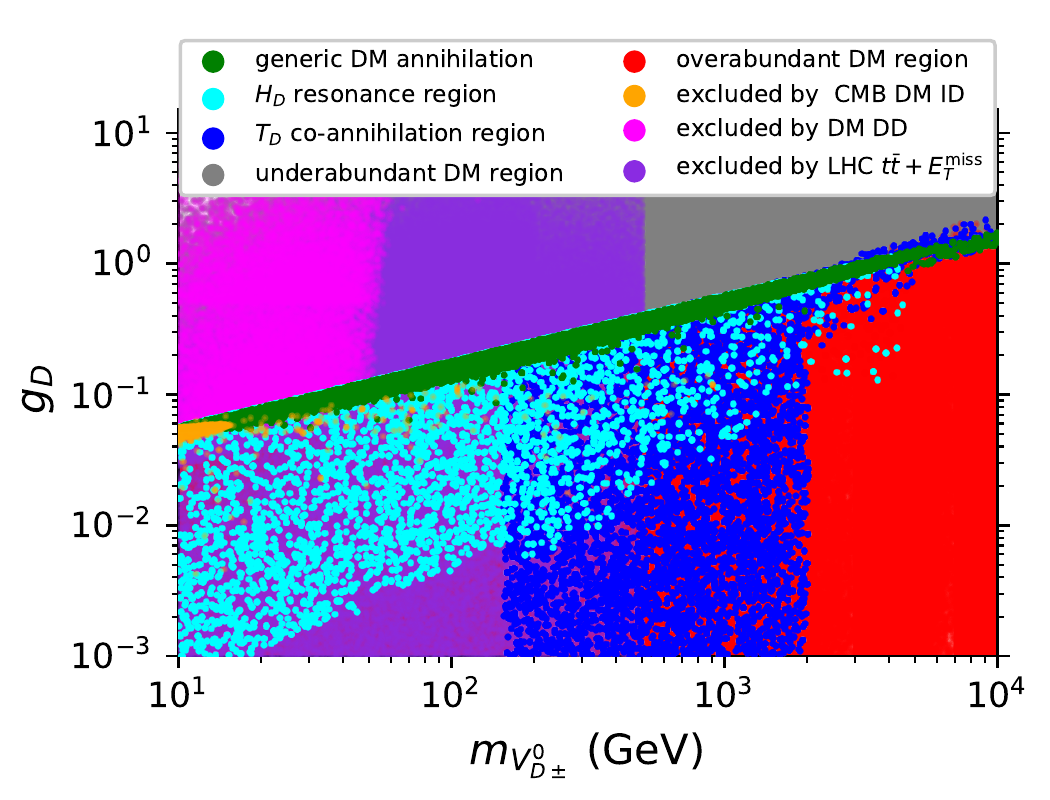}
\vskip -10pt
\caption{\label{fig:fullpicture} Excluded and allowed region of the parameter space of the model 
from the full five-dimensional scan ($\sin\theta_S=0$) of the parameter space projected into a ($g_{D},m_{V^0_{D\pm}}$) plane.}
\end{figure}

\begin{figure*}[htb]
\centering
\hspace*{-10pt}\includegraphics[width=.33\textwidth]{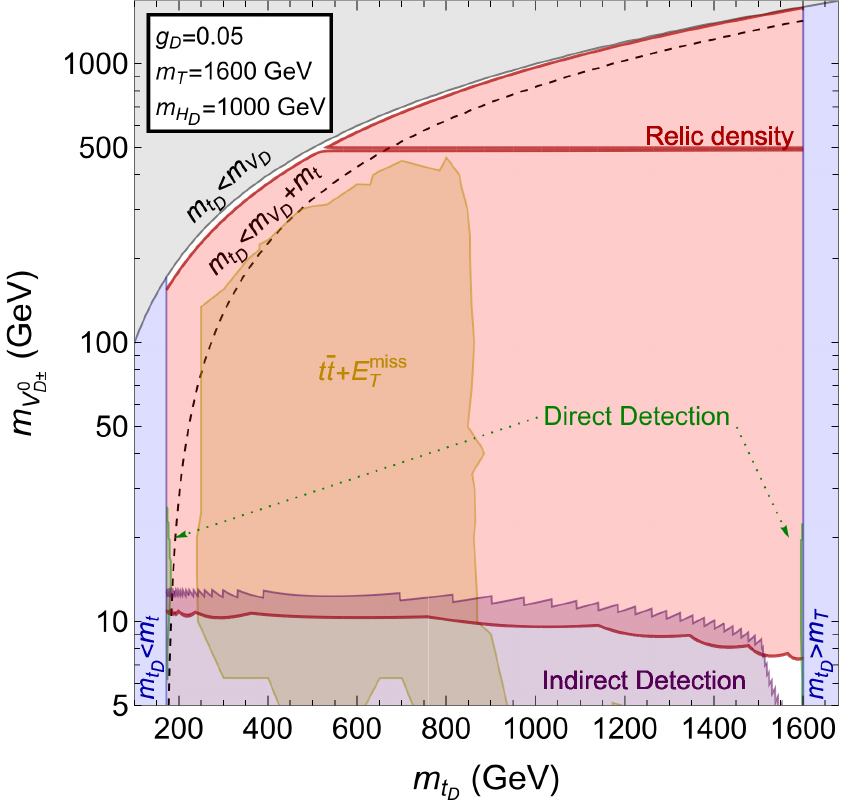}
\includegraphics[width=.33\textwidth]{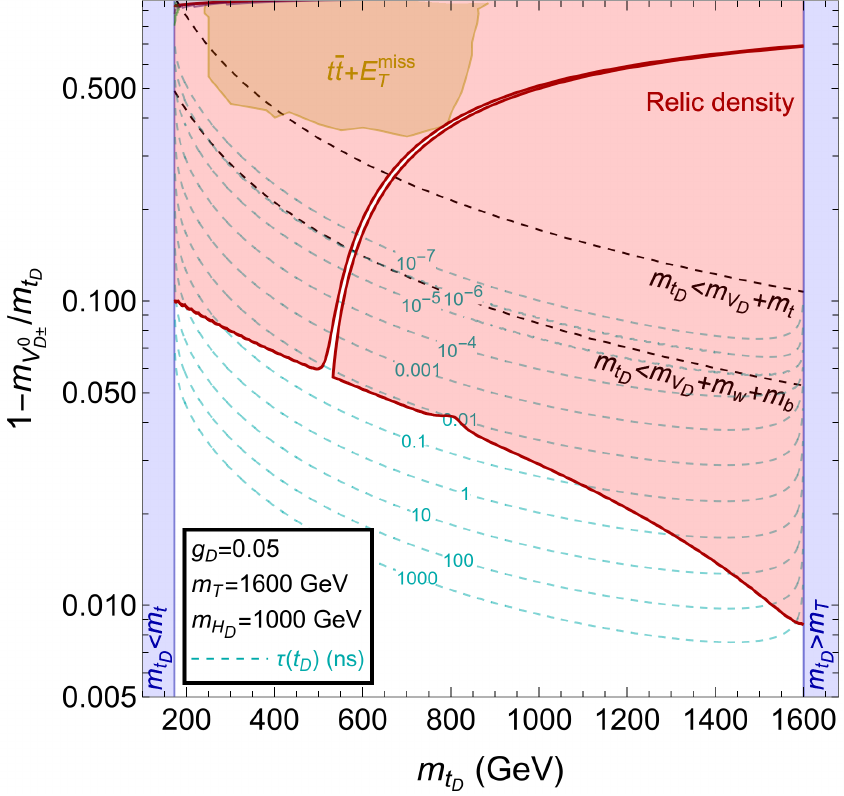}
\includegraphics[width=.33\textwidth]{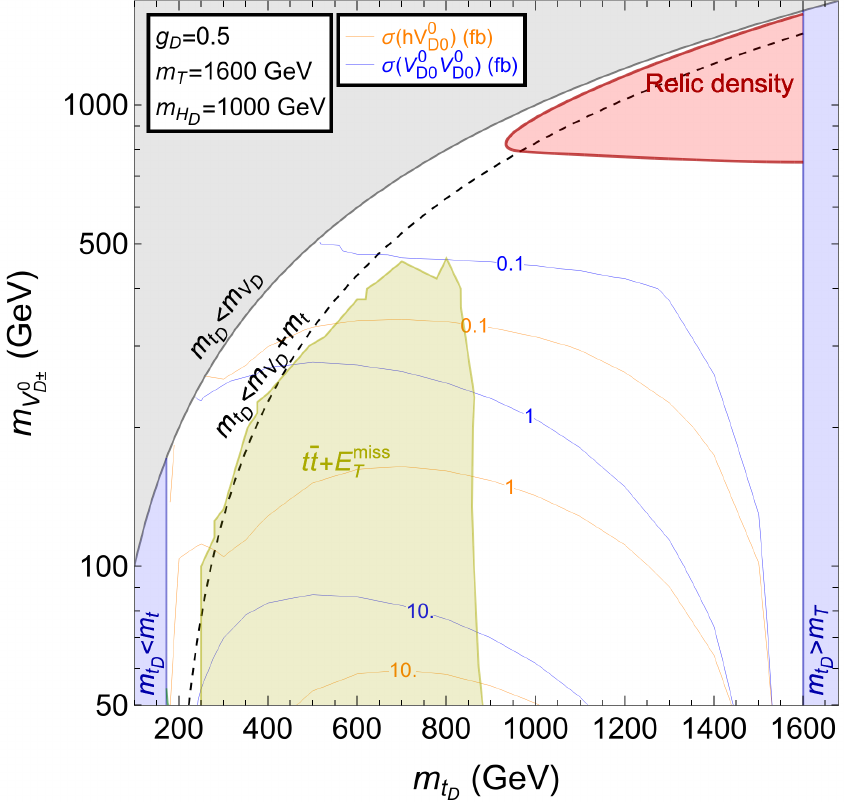}
\vskip -5pt
\caption{\label{fig:BPresults} Combination of constraints from LHC, relic density, DD and ID for the benchmark points in the $\{m_{t_{D}},m_{V^0_{D\pm}}\}$ (left and right panels) and $\{m_{t_{D}},1-{m_{V^0_{D\pm}}/m_{t_{D}}}\}$ (center panel) planes. The coloured regions are excluded. For relic density, the under-abundant region is considered as allowed and the borders of the excluded region correspond to the measured relic density value. Contours corresponding to different $t_{D}$ lifetimes are shown for the small mass splitting region.}
\end{figure*}

The complementarity of  cosmological and collider constraints has been studied by performing a comprehensive scan over the parameter space (excluding the fixed parameter $\sin\theta_S=0$) projected onto the ($g_{D},m_{V^0_{D\pm}}$) plane, shown in \cref{fig:fullpicture}. 
The allowed parameter space 
is indicated by the green, cyan and blue regions, presenting generic DM annihilation, dominated by the t-channel diagram of \cref{fig:feynmandiagrams}(a), resonance ($H_{D}$) and DM-$t_{D}$ co-annihilation regions, respectively, which satisfy  the relic density 
constraint from PLANCK within 5\%.
The generic DM annihilation determines a lower limit on $g_{D}$ as a function of  $m_{V^0_{D\pm}}$.
At the same time the $H_{D}$ resonant region allows to reduce $g_{D}$ values by up to two orders of magnitude, while the strong DM-$t_{D}$ co-annihilation channel allows for even lower values of $g_{D}$ for not so heavy DM. For $m_{V^0_{D\pm}}$ above 2\TeV, however, both the co-annihilation and resonant DM annihilation mechanisms are not effective anymore. Therefore, the region with low $g_{D}$ and large $m_{V^0_{D\pm}}$ has an over-abundant relic density, as indicated by the red colour. Notice also that the regions with low $m_{V^0_{D\pm}}$ and large $g_{D}$ values are partly excluded by DD and/or ID experiments as indicated by magenta and orange points, respectively. The region of DM masses which can be tested and excluded by the LHC is $m_{V^0_{D\pm}}\lesssim 400\GeV$, represented by the violet region.

To assess the relative role of the different constraints we identify representative benchmarks, characterised by different gauge couplings, $g_{D}=0.05$ and $g_{D}=0.5$, and fixed values for the masses, $m_T=1600\GeV$ and $m_{H_{D}}=1000\GeV$. For these points the gauge coupling is small enough to allow a perturbative treatment in a region of parameter space which can be tested by both collider and cosmological observables and the $\Z_2$-even partner of the top and  new scalar $H_{D}$ are heavy enough to easily evade current LHC bounds based on their decays into SM final states.
We show in \cref{fig:BPresults} the exclusion regions in the $\{m_{t_{D}},m_{V^0_{D\pm}}\}$ and $\{m_{t_{D}},1-{m_{V^0_{D\pm}}/m_{t_{D}}}\}$ planes to highlight the low $m_{V^0_{D\pm}}$ or low $m_{t_{D}}-m_{V^0_{D\pm}}$ regions, respectively. The masses of the DM candidate $V^0_{D\pm}$ and mediator $t_{D}$ are left as free parameters. 

The measured amount of relic density is satisfied only in specific regions: for $g_{D}=0.05$ most of the parameter space predicts an over-abundant relic density, except for an area where the mass difference between $t_{D}$ and the DM is less than $\sim10\%$ of the mediator mass (where $t_D$-$t_D$ and DM-$t_D$ co-annihilation processes dominate), a small area around $m_{V^0_{D\pm}}=m_{H_{D}}/2$ (DM annihilation via resonant $H_D$) and for $m_{V^0_{D\pm}}\lesssim10\GeV$. For larger values of $g_{D}$, annihilation processes become more effective, reducing the size of the excluded area in the lower $m_{V^0_{D\pm}}$ region, and eventually extending the under-abundant relic density region. 
The radiative mass split of the $V^0_{D0}$ and $V^0_{D\pm}$ bosons, $\Delta m_V=m_{V^0_{D\pm}} - m_{V^0_{D0}}$, due to fermion-loop mass corrections, plays a special role in the determination of 
relic density and ID rates.
The leading contribution to $\Delta m_V$ is determined by $T$ and $t_{D}$  loops to be 
\begin{equation}
\Delta m_V
=\frac{\epsilon^2 g_{D}^2 m_T^2}{32 \pi^2 m_{V^0_{D\pm}}} +o(\epsilon^2), \mbox{\ \ where\ \ } 
\epsilon=\frac{m_T^2-m_{t_{D}}^2}{m_T^2}.
\end{equation} 
Since $\Delta m_V>0$, the enhancement of the $V^0_{D+}V^0_{D-} \to V^0_{D0}V^0_{D0}$ process affects the relic density and ID signals. The complementarity of various constraints is especially evident for small values of $g_D$ in the low $m_{V^0_{D\pm}}$ region. {The region excluded by ID corresponds to small values of $m_{V^0_{D\pm}}$ for $g_D=0.05$, largely overlapping with the region excluded by relic density, and rapidly vanishes as $g_D$ increases.}
The excluded DD regions correspond to the limiting values $m_{t_D}\to m_t$ and $m_{t_D}\to m_T$ for low values of $m_{V^0_{D\pm}}$. In these regions the contribution of triangle loops becomes negligible with respect to the enhanced box diagrams (see \cref{fig:feynmandiagrams}(b)), which otherwise interfere destructively, reducing the cross-section of the DD process and evading the constraints.
The LHC bound is almost independent of the mass of $t_{D}$ until its mass difference with the DM reaches the top-quark threshold: in that region the $E_T^{\rm miss}$ decreases and the sensitivity of the CMS search reduces, allowing a small mass-gap region. As the process is QCD initiated, the bound is almost dependent on other parameters of the model. Processes of $V^0_{D0}$ pair production and associate production of $V^0_{D0}$ with the Higgs boson have cross-sections which scale with $g_D^4$ and $g_D^2$ respectively and are testable ($\sigma\gtrsim\mathcal{O}(10\fb)$) only for small ($\lesssim100$ GeV) values of $m_{V^0_{D0}}$ which 
partly excluded by ID constraints.
The model has an important feature, especially for small values of $g_{D}$ in the small DM-$t_D$ mass-gap region where the correct relic density is reproduced. 
In this region $t_D$ is long-lived (its lifetime in the small mass-gap region is shown in \cref{fig:BPresults}) and can be probed by dedicated searches at the LHC or future colliders.
Different $T$ or $H_{D}$ masses would not modify this qualitative picture.

This  minimal realisation of FPVDM has already great potential to explain DM phenomena together with several important implications for collider and non-collider DM searches.
Non-minimal FPVDM realisations would imply even richer sets of predictions and can be used to explain outstanding observed anomalies. For example, if the VL fermion interacts with the leptonic sector of the SM, new contributions might explain current lepton flavour anomalies~\cite{Crivellin:2021sff} or $(g-2)_\mu$~\cite{Muong-2:2021ojo} and at the same time would provide novel physics cases for future $e^+e^-$ colliders~\cite{Aicheler:2012bya,Baer:2013cma,An:2018dwb,FCC:2018evy}. 
Including mixing in the scalar sector, further VL partners or additional interactions of the same VL representation would open up a vast range of possibilities for future studies, both phenomenological and experimental, and would allow one to explore the complementarity between collider and non-collider observables in multiple scenarios.


\section*{Acknowledgements}

AB and SM acknowledge support from the STFC Consolidated Grant ST/L000296/1 and are partially
financed through the NExT Institute.
LP's work is supported by the Knut and Alice Wallenberg foundation under the SHIFT project, grant KAW 2017.0100. AD is grateful to the LABEX Lyon Institute of Origins (ANR-10-LABX-0066)  for its financial support within the program``Investissements d'Avenir". NT is supported by the scholarship from the Development and Promotion of Science
and Technology Talents Project (DPST). All authors
acknowledge the use of the IRIDIS High-Performance Computing Facility and associated
support services at the University of Southampton in completing this work.

\bibliography{bibliography}

\begin{thebibliography}{43}%
\makeatletter
\providecommand \@ifxundefined [1]{%
 \@ifx{#1\undefined}
}%
\providecommand \@ifnum [1]{%
 \ifnum #1\expandafter \@firstoftwo
 \else \expandafter \@secondoftwo
 \fi
}%
\providecommand \@ifx [1]{%
 \ifx #1\expandafter \@firstoftwo
 \else \expandafter \@secondoftwo
 \fi
}%
\providecommand \natexlab [1]{#1}%
\providecommand \enquote  [1]{``#1''}%
\providecommand \bibnamefont  [1]{#1}%
\providecommand \bibfnamefont [1]{#1}%
\providecommand \citenamefont [1]{#1}%
\providecommand \href@noop [0]{\@secondoftwo}%
\providecommand \href [0]{\begingroup \@sanitize@url \@href}%
\providecommand \@href[1]{\@@startlink{#1}\@@href}%
\providecommand \@@href[1]{\endgroup#1\@@endlink}%
\providecommand \@sanitize@url [0]{\catcode `\\12\catcode `\$12\catcode
  `\&12\catcode `\#12\catcode `\^12\catcode `\_12\catcode `\%12\relax}%
\providecommand \@@startlink[1]{}%
\providecommand \@@endlink[0]{}%
\providecommand \url  [0]{\begingroup\@sanitize@url \@url }%
\providecommand \@url [1]{\endgroup\@href {#1}{\urlprefix }}%
\providecommand \urlprefix  [0]{URL }%
\providecommand \Eprint [0]{\href }%
\providecommand \doibase [0]{https://doi.org/}%
\providecommand \selectlanguage [0]{\@gobble}%
\providecommand \bibinfo  [0]{\@secondoftwo}%
\providecommand \bibfield  [0]{\@secondoftwo}%
\providecommand \translation [1]{[#1]}%
\providecommand \BibitemOpen [0]{}%
\providecommand \bibitemStop [0]{}%
\providecommand \bibitemNoStop [0]{.\EOS\space}%
\providecommand \EOS [0]{\spacefactor3000\relax}%
\providecommand \BibitemShut  [1]{\csname bibitem#1\endcsname}%
\let\auto@bib@innerbib\@empty
\bibitem [{\citenamefont {Hambye}(2009)}]{Hambye:2008bq}%
  \BibitemOpen
  \bibfield  {author} {\bibinfo {author} {\bibfnamefont {T.}~\bibnamefont
  {Hambye}},\ }\bibfield  {title} {\bibinfo {title} {{Hidden vector dark
  matter}},\ }\href {https://doi.org/10.1088/1126-6708/2009/01/028} {\bibfield
  {journal} {\bibinfo  {journal} {JHEP}\ }\textbf {\bibinfo {volume} {01}},\
  \bibinfo {pages} {028}},\ \Eprint {https://arxiv.org/abs/0811.0172}
  {arXiv:0811.0172 [hep-ph]} \BibitemShut {NoStop}%
\bibitem [{\citenamefont {Chen}\ \emph {et~al.}(2009)\citenamefont {Chen},
  \citenamefont {Cline},\ and\ \citenamefont {Frey}}]{Chen:2009ab}%
  \BibitemOpen
  \bibfield  {author} {\bibinfo {author} {\bibfnamefont {F.}~\bibnamefont
  {Chen}}, \bibinfo {author} {\bibfnamefont {J.~M.}\ \bibnamefont {Cline}},\
  and\ \bibinfo {author} {\bibfnamefont {A.~R.}\ \bibnamefont {Frey}},\
  }\bibfield  {title} {\bibinfo {title} {{Nonabelian dark matter: Models and
  constraints}},\ }\href {https://doi.org/10.1103/PhysRevD.80.083516}
  {\bibfield  {journal} {\bibinfo  {journal} {Phys. Rev. D}\ }\textbf {\bibinfo
  {volume} {80}},\ \bibinfo {pages} {083516} (\bibinfo {year} {2009})},\
  \Eprint {https://arxiv.org/abs/0907.4746} {arXiv:0907.4746 [hep-ph]}
  \BibitemShut {NoStop}%
\bibitem [{\citenamefont {Diaz-Cruz}\ and\ \citenamefont
  {Ma}(2011)}]{DiazCruz:2010dc}%
  \BibitemOpen
  \bibfield  {author} {\bibinfo {author} {\bibfnamefont {J.}~\bibnamefont
  {Diaz-Cruz}}\ and\ \bibinfo {author} {\bibfnamefont {E.}~\bibnamefont {Ma}},\
  }\bibfield  {title} {\bibinfo {title} {{Neutral SU(2) Gauge Extension of the
  Standard Model and a Vector-Boson Dark-Matter Candidate}},\ }\href
  {https://doi.org/10.1016/j.physletb.2010.11.039} {\bibfield  {journal}
  {\bibinfo  {journal} {Phys. Lett. B}\ }\textbf {\bibinfo {volume} {695}},\
  \bibinfo {pages} {264} (\bibinfo {year} {2011})},\ \Eprint
  {https://arxiv.org/abs/1007.2631} {arXiv:1007.2631 [hep-ph]} \BibitemShut
  {NoStop}%
\bibitem [{\citenamefont {Bhattacharya}\ \emph {et~al.}(2012)\citenamefont
  {Bhattacharya}, \citenamefont {Diaz-Cruz}, \citenamefont {Ma},\ and\
  \citenamefont {Wegman}}]{Bhattacharya:2011tr}%
  \BibitemOpen
  \bibfield  {author} {\bibinfo {author} {\bibfnamefont {S.}~\bibnamefont
  {Bhattacharya}}, \bibinfo {author} {\bibfnamefont {J.}~\bibnamefont
  {Diaz-Cruz}}, \bibinfo {author} {\bibfnamefont {E.}~\bibnamefont {Ma}},\ and\
  \bibinfo {author} {\bibfnamefont {D.}~\bibnamefont {Wegman}},\ }\bibfield
  {title} {\bibinfo {title} {{Dark Vector-Gauge-Boson Model}},\ }\href
  {https://doi.org/10.1103/PhysRevD.85.055008} {\bibfield  {journal} {\bibinfo
  {journal} {Phys. Rev. D}\ }\textbf {\bibinfo {volume} {85}},\ \bibinfo
  {pages} {055008} (\bibinfo {year} {2012})},\ \Eprint
  {https://arxiv.org/abs/1107.2093} {arXiv:1107.2093 [hep-ph]} \BibitemShut
  {NoStop}%
\bibitem [{\citenamefont {Koorambas}(2013)}]{Koorambas:2013una}%
  \BibitemOpen
  \bibfield  {author} {\bibinfo {author} {\bibfnamefont {E.}~\bibnamefont
  {Koorambas}},\ }\bibfield  {title} {\bibinfo {title} {{Vector Gauge Boson
  Dark Matter for the $SU(N)$ Gauge Group Model}},\ }\href
  {https://doi.org/10.1007/s10773-013-1756-3} {\bibfield  {journal} {\bibinfo
  {journal} {Int. J. Theor. Phys.}\ }\textbf {\bibinfo {volume} {52}},\
  \bibinfo {pages} {4374} (\bibinfo {year} {2013})}\BibitemShut {NoStop}%
\bibitem [{\citenamefont {Fraser}\ \emph {et~al.}(2015)\citenamefont {Fraser},
  \citenamefont {Ma},\ and\ \citenamefont {Zakeri}}]{Fraser:2014yga}%
  \BibitemOpen
  \bibfield  {author} {\bibinfo {author} {\bibfnamefont {S.}~\bibnamefont
  {Fraser}}, \bibinfo {author} {\bibfnamefont {E.}~\bibnamefont {Ma}},\ and\
  \bibinfo {author} {\bibfnamefont {M.}~\bibnamefont {Zakeri}},\ }\bibfield
  {title} {\bibinfo {title} {{$SU(2)_N$ model of vector dark matter with a
  leptonic connection}},\ }\href {https://doi.org/10.1142/S0217751X15500189}
  {\bibfield  {journal} {\bibinfo  {journal} {Int. J. Mod. Phys. A}\ }\textbf
  {\bibinfo {volume} {30}},\ \bibinfo {pages} {1550018} (\bibinfo {year}
  {2015})},\ \Eprint {https://arxiv.org/abs/1409.1162} {arXiv:1409.1162
  [hep-ph]} \BibitemShut {NoStop}%
\bibitem [{\citenamefont {Hubisz}\ and\ \citenamefont
  {Meade}(2005)}]{Hubisz:2004ft}%
  \BibitemOpen
  \bibfield  {author} {\bibinfo {author} {\bibfnamefont {J.}~\bibnamefont
  {Hubisz}}\ and\ \bibinfo {author} {\bibfnamefont {P.}~\bibnamefont {Meade}},\
  }\bibfield  {title} {\bibinfo {title} {{Phenomenology of the littlest Higgs
  with T-parity}},\ }\href {https://doi.org/10.1103/PhysRevD.71.035016}
  {\bibfield  {journal} {\bibinfo  {journal} {Phys. Rev. D}\ }\textbf {\bibinfo
  {volume} {71}},\ \bibinfo {pages} {035016} (\bibinfo {year} {2005})},\
  \Eprint {https://arxiv.org/abs/hep-ph/0411264} {arXiv:hep-ph/0411264}
  \BibitemShut {NoStop}%
\bibitem [{\citenamefont {Huang}\ \emph {et~al.}(2016)\citenamefont {Huang},
  \citenamefont {Tsai},\ and\ \citenamefont {Yuan}}]{Huang:2015wts}%
  \BibitemOpen
  \bibfield  {author} {\bibinfo {author} {\bibfnamefont {W.-C.}\ \bibnamefont
  {Huang}}, \bibinfo {author} {\bibfnamefont {Y.-L.~S.}\ \bibnamefont {Tsai}},\
  and\ \bibinfo {author} {\bibfnamefont {T.-C.}\ \bibnamefont {Yuan}},\
  }\bibfield  {title} {\bibinfo {title} {{G2HDM : Gauged Two Higgs Doublet
  Model}},\ }\href {https://doi.org/10.1007/JHEP04(2016)019} {\bibfield
  {journal} {\bibinfo  {journal} {JHEP}\ }\textbf {\bibinfo {volume} {04}},\
  \bibinfo {pages} {019}},\ \Eprint {https://arxiv.org/abs/1512.00229}
  {arXiv:1512.00229 [hep-ph]} \BibitemShut {NoStop}%
\bibitem [{\citenamefont {Ko}\ and\ \citenamefont {Tang}(2017)}]{Ko:2016fcd}%
  \BibitemOpen
  \bibfield  {author} {\bibinfo {author} {\bibfnamefont {P.}~\bibnamefont
  {Ko}}\ and\ \bibinfo {author} {\bibfnamefont {Y.}~\bibnamefont {Tang}},\
  }\bibfield  {title} {\bibinfo {title} {{Residual Non-Abelian Dark Matter and
  Dark Radiation}},\ }\href {https://doi.org/10.1016/j.physletb.2017.02.033}
  {\bibfield  {journal} {\bibinfo  {journal} {Phys. Lett. B}\ }\textbf
  {\bibinfo {volume} {768}},\ \bibinfo {pages} {12} (\bibinfo {year} {2017})},\
  \Eprint {https://arxiv.org/abs/1609.02307} {arXiv:1609.02307 [hep-ph]}
  \BibitemShut {NoStop}%
\bibitem [{\citenamefont {Barman}\ \emph {et~al.}(2017)\citenamefont {Barman},
  \citenamefont {Bhattacharya}, \citenamefont {Patra},\ and\ \citenamefont
  {Chakrabortty}}]{Barman:2017yzr}%
  \BibitemOpen
  \bibfield  {author} {\bibinfo {author} {\bibfnamefont {B.}~\bibnamefont
  {Barman}}, \bibinfo {author} {\bibfnamefont {S.}~\bibnamefont
  {Bhattacharya}}, \bibinfo {author} {\bibfnamefont {S.~K.}\ \bibnamefont
  {Patra}},\ and\ \bibinfo {author} {\bibfnamefont {J.}~\bibnamefont
  {Chakrabortty}},\ }\bibfield  {title} {\bibinfo {title} {{Non-Abelian Vector
  Boson Dark Matter, its Unified Route and signatures at the LHC}},\ }\href
  {https://doi.org/10.1088/1475-7516/2017/12/021} {\bibfield  {journal}
  {\bibinfo  {journal} {JCAP}\ }\textbf {\bibinfo {volume} {12}},\ \bibinfo
  {pages} {021}},\ \Eprint {https://arxiv.org/abs/1704.04945} {arXiv:1704.04945
  [hep-ph]} \BibitemShut {NoStop}%
\bibitem [{\citenamefont {Huang}\ \emph {et~al.}(2018)\citenamefont {Huang},
  \citenamefont {Ishida}, \citenamefont {Lu}, \citenamefont {Tsai},\ and\
  \citenamefont {Yuan}}]{Huang:2017bto}%
  \BibitemOpen
  \bibfield  {author} {\bibinfo {author} {\bibfnamefont {W.-C.}\ \bibnamefont
  {Huang}}, \bibinfo {author} {\bibfnamefont {H.}~\bibnamefont {Ishida}},
  \bibinfo {author} {\bibfnamefont {C.-T.}\ \bibnamefont {Lu}}, \bibinfo
  {author} {\bibfnamefont {Y.-L.~S.}\ \bibnamefont {Tsai}},\ and\ \bibinfo
  {author} {\bibfnamefont {T.-C.}\ \bibnamefont {Yuan}},\ }\bibfield  {title}
  {\bibinfo {title} {{Signals of New Gauge Bosons in Gauged Two Higgs Doublet
  Model}},\ }\href {https://doi.org/10.1140/epjc/s10052-018-6067-7} {\bibfield
  {journal} {\bibinfo  {journal} {Eur. Phys. J. C}\ }\textbf {\bibinfo {volume}
  {78}},\ \bibinfo {pages} {613} (\bibinfo {year} {2018})},\ \Eprint
  {https://arxiv.org/abs/1708.02355} {arXiv:1708.02355 [hep-ph]} \BibitemShut
  {NoStop}%
\bibitem [{\citenamefont {Barman}\ \emph {et~al.}(2018)\citenamefont {Barman},
  \citenamefont {Bhattacharya},\ and\ \citenamefont {Zakeri}}]{Barman:2018esi}%
  \BibitemOpen
  \bibfield  {author} {\bibinfo {author} {\bibfnamefont {B.}~\bibnamefont
  {Barman}}, \bibinfo {author} {\bibfnamefont {S.}~\bibnamefont
  {Bhattacharya}},\ and\ \bibinfo {author} {\bibfnamefont {M.}~\bibnamefont
  {Zakeri}},\ }\bibfield  {title} {\bibinfo {title} {{Multipartite Dark Matter
  in $SU(2)_N$ extension of Standard Model and signatures at the LHC}},\ }\href
  {https://doi.org/10.1088/1475-7516/2018/09/023} {\bibfield  {journal}
  {\bibinfo  {journal} {JCAP}\ }\textbf {\bibinfo {volume} {09}},\ \bibinfo
  {pages} {023}},\ \Eprint {https://arxiv.org/abs/1806.01129} {arXiv:1806.01129
  [hep-ph]} \BibitemShut {NoStop}%
\bibitem [{\citenamefont {Barman}\ \emph {et~al.}(2020)\citenamefont {Barman},
  \citenamefont {Bhattacharya},\ and\ \citenamefont {Zakeri}}]{Barman:2019lvm}%
  \BibitemOpen
  \bibfield  {author} {\bibinfo {author} {\bibfnamefont {B.}~\bibnamefont
  {Barman}}, \bibinfo {author} {\bibfnamefont {S.}~\bibnamefont
  {Bhattacharya}},\ and\ \bibinfo {author} {\bibfnamefont {M.}~\bibnamefont
  {Zakeri}},\ }\bibfield  {title} {\bibinfo {title} {{Non-Abelian Vector Boson
  as FIMP Dark Matter}},\ }\href
  {https://doi.org/10.1088/1475-7516/2020/02/029} {\bibfield  {journal}
  {\bibinfo  {journal} {JCAP}\ }\textbf {\bibinfo {volume} {02}},\ \bibinfo
  {pages} {029}},\ \Eprint {https://arxiv.org/abs/1905.07236} {arXiv:1905.07236
  [hep-ph]} \BibitemShut {NoStop}%
\bibitem [{\citenamefont {Abe}\ \emph {et~al.}(2020)\citenamefont {Abe},
  \citenamefont {Fujiwara}, \citenamefont {Hisano},\ and\ \citenamefont
  {Matsushita}}]{Abe:2020mph}%
  \BibitemOpen
  \bibfield  {author} {\bibinfo {author} {\bibfnamefont {T.}~\bibnamefont
  {Abe}}, \bibinfo {author} {\bibfnamefont {M.}~\bibnamefont {Fujiwara}},
  \bibinfo {author} {\bibfnamefont {J.}~\bibnamefont {Hisano}},\ and\ \bibinfo
  {author} {\bibfnamefont {K.}~\bibnamefont {Matsushita}},\ }\bibfield  {title}
  {\bibinfo {title} {{A model of electroweakly interacting non-abelian vector
  dark matter}},\ }\href {https://doi.org/10.1007/JHEP07(2020)136} {\bibfield
  {journal} {\bibinfo  {journal} {JHEP}\ }\textbf {\bibinfo {volume} {07}},\
  \bibinfo {pages} {136}},\ \Eprint {https://arxiv.org/abs/2004.00884}
  {arXiv:2004.00884 [hep-ph]} \BibitemShut {NoStop}%
\bibitem [{\citenamefont {Chowdhury}\ and\ \citenamefont
  {Saad}(2021)}]{Chowdhury:2021tnm}%
  \BibitemOpen
  \bibfield  {author} {\bibinfo {author} {\bibfnamefont {T.~A.}\ \bibnamefont
  {Chowdhury}}\ and\ \bibinfo {author} {\bibfnamefont {S.}~\bibnamefont
  {Saad}},\ }\bibfield  {title} {\bibinfo {title} {{Non-Abelian vector dark
  matter and lepton g-2}},\ }\href
  {https://doi.org/10.1088/1475-7516/2021/10/014} {\bibfield  {journal}
  {\bibinfo  {journal} {JCAP}\ }\textbf {\bibinfo {volume} {10}},\ \bibinfo
  {pages} {014}},\ \Eprint {https://arxiv.org/abs/2107.11863} {arXiv:2107.11863
  [hep-ph]} \BibitemShut {NoStop}%
\bibitem [{\citenamefont {Baouche}\ \emph {et~al.}(2021)\citenamefont
  {Baouche}, \citenamefont {Ahriche}, \citenamefont {Faisel},\ and\
  \citenamefont {Nasri}}]{Baouche:2021wwa}%
  \BibitemOpen
  \bibfield  {author} {\bibinfo {author} {\bibfnamefont {N.}~\bibnamefont
  {Baouche}}, \bibinfo {author} {\bibfnamefont {A.}~\bibnamefont {Ahriche}},
  \bibinfo {author} {\bibfnamefont {G.}~\bibnamefont {Faisel}},\ and\ \bibinfo
  {author} {\bibfnamefont {S.}~\bibnamefont {Nasri}},\ }\bibfield  {title}
  {\bibinfo {title} {{Phenomenology of the hidden SU(2) vector dark matter
  model}},\ }\href {https://doi.org/10.1103/PhysRevD.104.075022} {\bibfield
  {journal} {\bibinfo  {journal} {Phys. Rev. D}\ }\textbf {\bibinfo {volume}
  {104}},\ \bibinfo {pages} {075022} (\bibinfo {year} {2021})},\ \Eprint
  {https://arxiv.org/abs/2105.14387} {arXiv:2105.14387 [hep-ph]} \BibitemShut
  {NoStop}%
\bibitem [{\citenamefont {Jim\'enez}\ \emph {et~al.}(2021)\citenamefont
  {Jim\'enez}, \citenamefont {Bettoni},\ and\ \citenamefont
  {Brax}}]{Jimenez:2020bgw}%
  \BibitemOpen
  \bibfield  {author} {\bibinfo {author} {\bibfnamefont {J.~B.}\ \bibnamefont
  {Jim\'enez}}, \bibinfo {author} {\bibfnamefont {D.}~\bibnamefont {Bettoni}},\
  and\ \bibinfo {author} {\bibfnamefont {P.}~\bibnamefont {Brax}},\ }\bibfield
  {title} {\bibinfo {title} {{Charged dark matter and the $H_0$ tension}},\
  }\href {https://doi.org/10.1103/PhysRevD.103.103505} {\bibfield  {journal}
  {\bibinfo  {journal} {Phys. Rev. D}\ }\textbf {\bibinfo {volume} {103}},\
  \bibinfo {pages} {103505} (\bibinfo {year} {2021})},\ \Eprint
  {https://arxiv.org/abs/2004.13677} {arXiv:2004.13677 [astro-ph.CO]}
  \BibitemShut {NoStop}%
\bibitem [{\citenamefont {Ma}\ and\ \citenamefont
  {Cacciapaglia}(2016)}]{Ma:2015gra}%
  \BibitemOpen
  \bibfield  {author} {\bibinfo {author} {\bibfnamefont {T.}~\bibnamefont
  {Ma}}\ and\ \bibinfo {author} {\bibfnamefont {G.}~\bibnamefont
  {Cacciapaglia}},\ }\bibfield  {title} {\bibinfo {title} {{Fundamental
  Composite 2HDM: SU(N) with 4 flavours}},\ }\href
  {https://doi.org/10.1007/JHEP03(2016)211} {\bibfield  {journal} {\bibinfo
  {journal} {JHEP}\ }\textbf {\bibinfo {volume} {03}},\ \bibinfo {pages}
  {211}},\ \Eprint {https://arxiv.org/abs/1508.07014} {arXiv:1508.07014
  [hep-ph]} \BibitemShut {NoStop}%
\bibitem [{\citenamefont {Wu}\ \emph {et~al.}(2017)\citenamefont {Wu},
  \citenamefont {Ma}, \citenamefont {Zhang},\ and\ \citenamefont
  {Cacciapaglia}}]{Wu:2017iji}%
  \BibitemOpen
  \bibfield  {author} {\bibinfo {author} {\bibfnamefont {Y.}~\bibnamefont
  {Wu}}, \bibinfo {author} {\bibfnamefont {T.}~\bibnamefont {Ma}}, \bibinfo
  {author} {\bibfnamefont {B.}~\bibnamefont {Zhang}},\ and\ \bibinfo {author}
  {\bibfnamefont {G.}~\bibnamefont {Cacciapaglia}},\ }\bibfield  {title}
  {\bibinfo {title} {{Composite Dark Matter and Higgs}},\ }\href
  {https://doi.org/10.1007/JHEP11(2017)058} {\bibfield  {journal} {\bibinfo
  {journal} {JHEP}\ }\textbf {\bibinfo {volume} {11}},\ \bibinfo {pages}
  {058}},\ \Eprint {https://arxiv.org/abs/1703.06903} {arXiv:1703.06903
  [hep-ph]} \BibitemShut {NoStop}%
\bibitem [{\citenamefont {Baek}\ \emph {et~al.}(2018)\citenamefont {Baek},
  \citenamefont {Ko},\ and\ \citenamefont {Wu}}]{Baek:2017ykw}%
  \BibitemOpen
  \bibfield  {author} {\bibinfo {author} {\bibfnamefont {S.}~\bibnamefont
  {Baek}}, \bibinfo {author} {\bibfnamefont {P.}~\bibnamefont {Ko}},\ and\
  \bibinfo {author} {\bibfnamefont {P.}~\bibnamefont {Wu}},\ }\bibfield
  {title} {\bibinfo {title} {{Heavy quark-philic scalar dark matter with a
  vector-like fermion portal}},\ }\href
  {https://doi.org/10.1088/1475-7516/2018/07/008} {\bibfield  {journal}
  {\bibinfo  {journal} {JCAP}\ }\textbf {\bibinfo {volume} {07}},\ \bibinfo
  {pages} {008}},\ \Eprint {https://arxiv.org/abs/1709.00697} {arXiv:1709.00697
  [hep-ph]} \BibitemShut {NoStop}%
\bibitem [{\citenamefont {Colucci}\ \emph {et~al.}(2018)\citenamefont
  {Colucci}, \citenamefont {Fuks}, \citenamefont {Giacchino}, \citenamefont
  {Lopez~Honorez}, \citenamefont {Tytgat},\ and\ \citenamefont
  {Vandecasteele}}]{Colucci:2018vxz}%
  \BibitemOpen
  \bibfield  {author} {\bibinfo {author} {\bibfnamefont {S.}~\bibnamefont
  {Colucci}}, \bibinfo {author} {\bibfnamefont {B.}~\bibnamefont {Fuks}},
  \bibinfo {author} {\bibfnamefont {F.}~\bibnamefont {Giacchino}}, \bibinfo
  {author} {\bibfnamefont {L.}~\bibnamefont {Lopez~Honorez}}, \bibinfo {author}
  {\bibfnamefont {M.~H.~G.}\ \bibnamefont {Tytgat}},\ and\ \bibinfo {author}
  {\bibfnamefont {J.}~\bibnamefont {Vandecasteele}},\ }\bibfield  {title}
  {\bibinfo {title} {{Top-philic Vector-Like Portal to Scalar Dark Matter}},\
  }\href {https://doi.org/10.1103/PhysRevD.98.035002} {\bibfield  {journal}
  {\bibinfo  {journal} {Phys. Rev. D}\ }\textbf {\bibinfo {volume} {98}},\
  \bibinfo {pages} {035002} (\bibinfo {year} {2018})},\ \Eprint
  {https://arxiv.org/abs/1804.05068} {arXiv:1804.05068 [hep-ph]} \BibitemShut
  {NoStop}%
\bibitem [{\citenamefont {Semenov}(2009)}]{Semenov2009}%
  \BibitemOpen
  \bibfield  {author} {\bibinfo {author} {\bibfnamefont {A.}~\bibnamefont
  {Semenov}},\ }\bibfield  {title} {\bibinfo {title} {Lanhep—a package for
  the automatic generation of feynman rules in field theory. version 3.0},\
  }\href {https://doi.org/10.1016/j.cpc.2008.10.012} {\bibfield  {journal}
  {\bibinfo  {journal} {Computer Physics Communications}\ }\textbf {\bibinfo
  {volume} {180}},\ \bibinfo {pages} {431–454} (\bibinfo {year}
  {2009})}\BibitemShut {NoStop}%
\bibitem [{\citenamefont {Alloul}\ \emph {et~al.}(2014)\citenamefont {Alloul},
  \citenamefont {Christensen}, \citenamefont {Degrande}, \citenamefont {Duhr},\
  and\ \citenamefont {Fuks}}]{Alloul:2013bka}%
  \BibitemOpen
  \bibfield  {author} {\bibinfo {author} {\bibfnamefont {A.}~\bibnamefont
  {Alloul}}, \bibinfo {author} {\bibfnamefont {N.~D.}\ \bibnamefont
  {Christensen}}, \bibinfo {author} {\bibfnamefont {C.}~\bibnamefont
  {Degrande}}, \bibinfo {author} {\bibfnamefont {C.}~\bibnamefont {Duhr}},\
  and\ \bibinfo {author} {\bibfnamefont {B.}~\bibnamefont {Fuks}},\ }\bibfield
  {title} {\bibinfo {title} {{FeynRules 2.0 - A complete toolbox for tree-level
  phenomenology}},\ }\href {https://doi.org/10.1016/j.cpc.2014.04.012}
  {\bibfield  {journal} {\bibinfo  {journal} {Comput. Phys. Commun.}\ }\textbf
  {\bibinfo {volume} {185}},\ \bibinfo {pages} {2250} (\bibinfo {year}
  {2014})},\ \Eprint {https://arxiv.org/abs/1310.1921} {arXiv:1310.1921
  [hep-ph]} \BibitemShut {NoStop}%
\bibitem [{\citenamefont {Belyaev}\ \emph {et~al.}(2013)\citenamefont
  {Belyaev}, \citenamefont {Christensen},\ and\ \citenamefont
  {Pukhov}}]{Belyaev:2012qa}%
  \BibitemOpen
  \bibfield  {author} {\bibinfo {author} {\bibfnamefont {A.}~\bibnamefont
  {Belyaev}}, \bibinfo {author} {\bibfnamefont {N.~D.}\ \bibnamefont
  {Christensen}},\ and\ \bibinfo {author} {\bibfnamefont {A.}~\bibnamefont
  {Pukhov}},\ }\bibfield  {title} {\bibinfo {title} {{CalcHEP 3.4 for collider
  physics within and beyond the Standard Model}},\ }\href
  {https://doi.org/10.1016/j.cpc.2013.01.014} {\bibfield  {journal} {\bibinfo
  {journal} {Comput. Phys. Commun.}\ }\textbf {\bibinfo {volume} {184}},\
  \bibinfo {pages} {1729} (\bibinfo {year} {2013})},\ \Eprint
  {https://arxiv.org/abs/1207.6082} {arXiv:1207.6082 [hep-ph]} \BibitemShut
  {NoStop}%
\bibitem [{\citenamefont {Degrande}\ \emph {et~al.}(2012)\citenamefont
  {Degrande}, \citenamefont {Duhr}, \citenamefont {Fuks}, \citenamefont
  {Grellscheid}, \citenamefont {Mattelaer},\ and\ \citenamefont
  {Reiter}}]{Degrande:2011ua}%
  \BibitemOpen
  \bibfield  {author} {\bibinfo {author} {\bibfnamefont {C.}~\bibnamefont
  {Degrande}}, \bibinfo {author} {\bibfnamefont {C.}~\bibnamefont {Duhr}},
  \bibinfo {author} {\bibfnamefont {B.}~\bibnamefont {Fuks}}, \bibinfo {author}
  {\bibfnamefont {D.}~\bibnamefont {Grellscheid}}, \bibinfo {author}
  {\bibfnamefont {O.}~\bibnamefont {Mattelaer}},\ and\ \bibinfo {author}
  {\bibfnamefont {T.}~\bibnamefont {Reiter}},\ }\bibfield  {title} {\bibinfo
  {title} {{UFO - The Universal FeynRules Output}},\ }\href
  {https://doi.org/10.1016/j.cpc.2012.01.022} {\bibfield  {journal} {\bibinfo
  {journal} {Comput. Phys. Commun.}\ }\textbf {\bibinfo {volume} {183}},\
  \bibinfo {pages} {1201} (\bibinfo {year} {2012})},\ \Eprint
  {https://arxiv.org/abs/1108.2040} {arXiv:1108.2040 [hep-ph]} \BibitemShut
  {NoStop}%
\bibitem [{\citenamefont {Hahn}(2001)}]{Hahn:2000kx}%
  \BibitemOpen
  \bibfield  {author} {\bibinfo {author} {\bibfnamefont {T.}~\bibnamefont
  {Hahn}},\ }\bibfield  {title} {\bibinfo {title} {{Generating Feynman diagrams
  and amplitudes with FeynArts 3}},\ }\href
  {https://doi.org/10.1016/S0010-4655(01)00290-9} {\bibfield  {journal}
  {\bibinfo  {journal} {Comput. Phys. Commun.}\ }\textbf {\bibinfo {volume}
  {140}},\ \bibinfo {pages} {418} (\bibinfo {year} {2001})},\ \Eprint
  {https://arxiv.org/abs/hep-ph/0012260} {arXiv:hep-ph/0012260} \BibitemShut
  {NoStop}%
\bibitem [{\citenamefont {Bondarenko}\ \emph {et~al.}(2012)\citenamefont
  {Bondarenko}, \citenamefont {Belyaev}, \citenamefont {Blandford},
  \citenamefont {Basso}, \citenamefont {Boos}, \citenamefont {Bunichev} \emph
  {et~al.}}]{hepmdb}%
  \BibitemOpen
  \bibfield  {author} {\bibinfo {author} {\bibfnamefont {M.}~\bibnamefont
  {Bondarenko}}, \bibinfo {author} {\bibfnamefont {A.}~\bibnamefont {Belyaev}},
  \bibinfo {author} {\bibfnamefont {J.}~\bibnamefont {Blandford}}, \bibinfo
  {author} {\bibfnamefont {L.}~\bibnamefont {Basso}}, \bibinfo {author}
  {\bibfnamefont {E.}~\bibnamefont {Boos}}, \bibinfo {author} {\bibfnamefont
  {V.}~\bibnamefont {Bunichev}}, \emph {et~al.},\ }\bibfield  {title} {\bibinfo
  {title} {{High Energy Physics Model Database : Towards decoding of the
  underlying theory (within Les Houches 2011: Physics at TeV Colliders New
  Physics Working Group Report)}},\ }\href {https://hepmdb.soton.ac.uk} {\
  (\bibinfo {year} {2012})},\ \Eprint {https://arxiv.org/abs/1203.1488}
  {arXiv:1203.1488 [hep-ph]} \BibitemShut {NoStop}%
\bibitem [{\citenamefont {Belanger}\ \emph {et~al.}(2021)\citenamefont
  {Belanger}, \citenamefont {Mjallal},\ and\ \citenamefont
  {Pukhov}}]{Belanger:2020gnr}%
  \BibitemOpen
  \bibfield  {author} {\bibinfo {author} {\bibfnamefont {G.}~\bibnamefont
  {Belanger}}, \bibinfo {author} {\bibfnamefont {A.}~\bibnamefont {Mjallal}},\
  and\ \bibinfo {author} {\bibfnamefont {A.}~\bibnamefont {Pukhov}},\
  }\bibfield  {title} {\bibinfo {title} {{Recasting direct detection limits
  within micrOMEGAs and implication for non-standard Dark Matter scenarios}},\
  }\href {https://doi.org/10.1140/epjc/s10052-021-09012-z} {\bibfield
  {journal} {\bibinfo  {journal} {Eur. Phys. J. C}\ }\textbf {\bibinfo {volume}
  {81}},\ \bibinfo {pages} {239} (\bibinfo {year} {2021})},\ \Eprint
  {https://arxiv.org/abs/2003.08621} {arXiv:2003.08621 [hep-ph]} \BibitemShut
  {NoStop}%
\bibitem [{\citenamefont {Alwall}\ \emph {et~al.}(2014)\citenamefont {Alwall},
  \citenamefont {Frederix}, \citenamefont {Frixione}, \citenamefont {Hirschi},
  \citenamefont {Maltoni}, \citenamefont {Mattelaer}, \citenamefont {Shao},
  \citenamefont {Stelzer}, \citenamefont {Torrielli},\ and\ \citenamefont
  {Zaro}}]{Alwall:2014hca}%
  \BibitemOpen
  \bibfield  {author} {\bibinfo {author} {\bibfnamefont {J.}~\bibnamefont
  {Alwall}}, \bibinfo {author} {\bibfnamefont {R.}~\bibnamefont {Frederix}},
  \bibinfo {author} {\bibfnamefont {S.}~\bibnamefont {Frixione}}, \bibinfo
  {author} {\bibfnamefont {V.}~\bibnamefont {Hirschi}}, \bibinfo {author}
  {\bibfnamefont {F.}~\bibnamefont {Maltoni}}, \bibinfo {author} {\bibfnamefont
  {O.}~\bibnamefont {Mattelaer}}, \bibinfo {author} {\bibfnamefont {H.~S.}\
  \bibnamefont {Shao}}, \bibinfo {author} {\bibfnamefont {T.}~\bibnamefont
  {Stelzer}}, \bibinfo {author} {\bibfnamefont {P.}~\bibnamefont {Torrielli}},\
  and\ \bibinfo {author} {\bibfnamefont {M.}~\bibnamefont {Zaro}},\ }\bibfield
  {title} {\bibinfo {title} {{The automated computation of tree-level and
  next-to-leading order differential cross sections, and their matching to
  parton shower simulations}},\ }\href
  {https://doi.org/10.1007/JHEP07(2014)079} {\bibfield  {journal} {\bibinfo
  {journal} {JHEP}\ }\textbf {\bibinfo {volume} {07}},\ \bibinfo {pages}
  {079}},\ \Eprint {https://arxiv.org/abs/1405.0301} {arXiv:1405.0301 [hep-ph]}
  \BibitemShut {NoStop}%
\bibitem [{\citenamefont {Ball}\ \emph {et~al.}(2015)\citenamefont {Ball} \emph
  {et~al.}}]{NNPDF:2014otw}%
  \BibitemOpen
  \bibfield  {author} {\bibinfo {author} {\bibfnamefont {R.~D.}\ \bibnamefont
  {Ball}} \emph {et~al.} (\bibinfo {collaboration} {NNPDF}),\ }\bibfield
  {title} {\bibinfo {title} {{Parton distributions for the LHC Run II}},\
  }\href {https://doi.org/10.1007/JHEP04(2015)040} {\bibfield  {journal}
  {\bibinfo  {journal} {JHEP}\ }\textbf {\bibinfo {volume} {04}},\ \bibinfo
  {pages} {040}},\ \Eprint {https://arxiv.org/abs/1410.8849} {arXiv:1410.8849
  [hep-ph]} \BibitemShut {NoStop}%
\bibitem [{\citenamefont {Buckley}\ \emph {et~al.}(2015)\citenamefont
  {Buckley}, \citenamefont {Ferrando}, \citenamefont {Lloyd}, \citenamefont
  {Nordstr\"om}, \citenamefont {Page}, \citenamefont {R\"ufenacht},
  \citenamefont {Sch\"onherr},\ and\ \citenamefont {Watt}}]{Buckley:2014ana}%
  \BibitemOpen
  \bibfield  {author} {\bibinfo {author} {\bibfnamefont {A.}~\bibnamefont
  {Buckley}}, \bibinfo {author} {\bibfnamefont {J.}~\bibnamefont {Ferrando}},
  \bibinfo {author} {\bibfnamefont {S.}~\bibnamefont {Lloyd}}, \bibinfo
  {author} {\bibfnamefont {K.}~\bibnamefont {Nordstr\"om}}, \bibinfo {author}
  {\bibfnamefont {B.}~\bibnamefont {Page}}, \bibinfo {author} {\bibfnamefont
  {M.}~\bibnamefont {R\"ufenacht}}, \bibinfo {author} {\bibfnamefont
  {M.}~\bibnamefont {Sch\"onherr}},\ and\ \bibinfo {author} {\bibfnamefont
  {G.}~\bibnamefont {Watt}},\ }\bibfield  {title} {\bibinfo {title} {{LHAPDF6:
  parton density access in the LHC precision era}},\ }\href
  {https://doi.org/10.1140/epjc/s10052-015-3318-8} {\bibfield  {journal}
  {\bibinfo  {journal} {Eur. Phys. J. C}\ }\textbf {\bibinfo {volume} {75}},\
  \bibinfo {pages} {132} (\bibinfo {year} {2015})},\ \Eprint
  {https://arxiv.org/abs/1412.7420} {arXiv:1412.7420 [hep-ph]} \BibitemShut
  {NoStop}%
\bibitem [{\citenamefont {Hahn}\ \emph {et~al.}(2016)\citenamefont {Hahn},
  \citenamefont {Pa\ss{}ehr},\ and\ \citenamefont
  {Schappacher}}]{Hahn:2016ebn}%
  \BibitemOpen
  \bibfield  {author} {\bibinfo {author} {\bibfnamefont {T.}~\bibnamefont
  {Hahn}}, \bibinfo {author} {\bibfnamefont {S.}~\bibnamefont {Pa\ss{}ehr}},\
  and\ \bibinfo {author} {\bibfnamefont {C.}~\bibnamefont {Schappacher}},\
  }\bibfield  {title} {\bibinfo {title} {{FormCalc 9 and Extensions}},\ }\href
  {https://doi.org/10.1088/1742-6596/762/1/012065} {\bibfield  {journal}
  {\bibinfo  {journal} {PoS}\ }\textbf {\bibinfo {volume} {LL2016}},\ \bibinfo
  {pages} {068} (\bibinfo {year} {2016})},\ \Eprint
  {https://arxiv.org/abs/1604.04611} {arXiv:1604.04611 [hep-ph]} \BibitemShut
  {NoStop}%
\bibitem [{\citenamefont {Aghanim}\ \emph {et~al.}(2020)\citenamefont {Aghanim}
  \emph {et~al.}}]{Planck:2018vyg}%
  \BibitemOpen
  \bibfield  {author} {\bibinfo {author} {\bibfnamefont {N.}~\bibnamefont
  {Aghanim}} \emph {et~al.} (\bibinfo {collaboration} {Planck}),\ }\bibfield
  {title} {\bibinfo {title} {{Planck 2018 results. VI. Cosmological
  parameters}},\ }\href {https://doi.org/10.1051/0004-6361/201833910}
  {\bibfield  {journal} {\bibinfo  {journal} {Astron. Astrophys.}\ }\textbf
  {\bibinfo {volume} {641}},\ \bibinfo {pages} {A6} (\bibinfo {year} {2020})},\
  \bibinfo {note} {[Erratum: Astron.Astrophys. 652, C4 (2021)]},\ \Eprint
  {https://arxiv.org/abs/1807.06209} {arXiv:1807.06209 [astro-ph.CO]}
  \BibitemShut {NoStop}%
\bibitem [{\citenamefont {Aprile}\ \emph {et~al.}(2018)\citenamefont {Aprile}
  \emph {et~al.}}]{XENON:2018voc}%
  \BibitemOpen
  \bibfield  {author} {\bibinfo {author} {\bibfnamefont {E.}~\bibnamefont
  {Aprile}} \emph {et~al.} (\bibinfo {collaboration} {XENON}),\ }\bibfield
  {title} {\bibinfo {title} {{Dark Matter Search Results from a One Ton-Year
  Exposure of XENON1T}},\ }\href
  {https://doi.org/10.1103/PhysRevLett.121.111302} {\bibfield  {journal}
  {\bibinfo  {journal} {Phys. Rev. Lett.}\ }\textbf {\bibinfo {volume} {121}},\
  \bibinfo {pages} {111302} (\bibinfo {year} {2018})},\ \Eprint
  {https://arxiv.org/abs/1805.12562} {arXiv:1805.12562 [astro-ph.CO]}
  \BibitemShut {NoStop}%
\bibitem [{\citenamefont {Sirunyan}\ \emph {et~al.}(2018)\citenamefont
  {Sirunyan} \emph {et~al.}}]{CMS:2017jrd}%
  \BibitemOpen
  \bibfield  {author} {\bibinfo {author} {\bibfnamefont {A.~M.}\ \bibnamefont
  {Sirunyan}} \emph {et~al.} (\bibinfo {collaboration} {CMS}),\ }\bibfield
  {title} {\bibinfo {title} {{Search for top squarks and dark matter particles
  in opposite-charge dilepton final states at $\sqrt{s}=$ 13 TeV}},\ }\href
  {https://doi.org/10.1103/PhysRevD.97.032009} {\bibfield  {journal} {\bibinfo
  {journal} {Phys. Rev. D}\ }\textbf {\bibinfo {volume} {97}},\ \bibinfo
  {pages} {032009} (\bibinfo {year} {2018})},\ \Eprint
  {https://arxiv.org/abs/1711.00752} {arXiv:1711.00752 [hep-ex]} \BibitemShut
  {NoStop}%
\bibitem [{\citenamefont {Conte}\ and\ \citenamefont
  {Fuks}(2018)}]{Conte:2018vmg}%
  \BibitemOpen
  \bibfield  {author} {\bibinfo {author} {\bibfnamefont {E.}~\bibnamefont
  {Conte}}\ and\ \bibinfo {author} {\bibfnamefont {B.}~\bibnamefont {Fuks}},\
  }\bibfield  {title} {\bibinfo {title} {{Confronting new physics theories to
  LHC data with MADANALYSIS 5}},\ }\href
  {https://doi.org/10.1142/S0217751X18300272} {\bibfield  {journal} {\bibinfo
  {journal} {Int. J. Mod. Phys. A}\ }\textbf {\bibinfo {volume} {33}},\
  \bibinfo {pages} {1830027} (\bibinfo {year} {2018})},\ \Eprint
  {https://arxiv.org/abs/1808.00480} {arXiv:1808.00480 [hep-ph]} \BibitemShut
  {NoStop}%
\bibitem [{\citenamefont {Bein}\ \emph {et~al.}(2021)\citenamefont {Bein},
  \citenamefont {Choi}, \citenamefont {Fuks}, \citenamefont {Jeong},
  \citenamefont {Kang}, \citenamefont {Li},\ and\ \citenamefont
  {Sonneveld}}]{DVN/BQM0T3_2021}%
  \BibitemOpen
  \bibfield  {author} {\bibinfo {author} {\bibfnamefont {S.}~\bibnamefont
  {Bein}}, \bibinfo {author} {\bibfnamefont {S.-M.}\ \bibnamefont {Choi}},
  \bibinfo {author} {\bibfnamefont {B.}~\bibnamefont {Fuks}}, \bibinfo {author}
  {\bibfnamefont {S.}~\bibnamefont {Jeong}}, \bibinfo {author} {\bibfnamefont
  {D.~W.}\ \bibnamefont {Kang}}, \bibinfo {author} {\bibfnamefont
  {J.}~\bibnamefont {Li}},\ and\ \bibinfo {author} {\bibfnamefont
  {J.}~\bibnamefont {Sonneveld}},\ }\href {https://doi.org/10.14428/DVN/BQM0T3}
  {\bibinfo {title} {{Implementation of a search for stops in the di-lepton +
  missing energy channel (35.9 fb-1; 13 TeV; CMS-SUS-17-001)}}} (\bibinfo
  {year} {2021})\BibitemShut {NoStop}%
\bibitem [{\citenamefont {Crivellin}\ and\ \citenamefont
  {Hoferichter}(2021)}]{Crivellin:2021sff}%
  \BibitemOpen
  \bibfield  {author} {\bibinfo {author} {\bibfnamefont {A.}~\bibnamefont
  {Crivellin}}\ and\ \bibinfo {author} {\bibfnamefont {M.}~\bibnamefont
  {Hoferichter}},\ }\bibfield  {title} {\bibinfo {title} {{Hints of lepton
  flavor universality violations}},\ }\href
  {https://doi.org/10.1126/science.abk2450} {\bibfield  {journal} {\bibinfo
  {journal} {Science}\ }\textbf {\bibinfo {volume} {374}},\ \bibinfo {pages}
  {1051} (\bibinfo {year} {2021})},\ \Eprint {https://arxiv.org/abs/2111.12739}
  {arXiv:2111.12739 [hep-ph]} \BibitemShut {NoStop}%
\bibitem [{\citenamefont {Abi}\ \emph {et~al.}(2021)\citenamefont {Abi} \emph
  {et~al.}}]{Muong-2:2021ojo}%
  \BibitemOpen
  \bibfield  {author} {\bibinfo {author} {\bibfnamefont {B.}~\bibnamefont
  {Abi}} \emph {et~al.} (\bibinfo {collaboration} {Muon g-2}),\ }\bibfield
  {title} {\bibinfo {title} {{Measurement of the Positive Muon Anomalous
  Magnetic Moment to 0.46 ppm}},\ }\href
  {https://doi.org/10.1103/PhysRevLett.126.141801} {\bibfield  {journal}
  {\bibinfo  {journal} {Phys. Rev. Lett.}\ }\textbf {\bibinfo {volume} {126}},\
  \bibinfo {pages} {141801} (\bibinfo {year} {2021})},\ \Eprint
  {https://arxiv.org/abs/2104.03281} {arXiv:2104.03281 [hep-ex]} \BibitemShut
  {NoStop}%
\bibitem [{Aic(2012)}]{Aicheler:2012bya}%
  \BibitemOpen
  \bibfield  {title} {\bibinfo {title} {{A Multi-TeV Linear Collider Based on
  CLIC Technology}: {CLIC Conceptual Design Report}}\ }\href
  {https://doi.org/10.5170/CERN-2012-007} {10.5170/CERN-2012-007} (\bibinfo
  {year} {2012})\BibitemShut {NoStop}%
\bibitem [{Bae(2013)}]{Baer:2013cma}%
  \BibitemOpen
  \bibfield  {title} {\bibinfo {title} {{The International Linear Collider
  Technical Design Report - Volume 2: Physics}},\ }\href@noop {} {\  (\bibinfo
  {year} {2013})},\ \Eprint {https://arxiv.org/abs/1306.6352} {arXiv:1306.6352
  [hep-ph]} \BibitemShut {NoStop}%
\bibitem [{\citenamefont {An}\ \emph {et~al.}(2019)\citenamefont {An} \emph
  {et~al.}}]{An:2018dwb}%
  \BibitemOpen
  \bibfield  {author} {\bibinfo {author} {\bibfnamefont {F.}~\bibnamefont {An}}
  \emph {et~al.},\ }\bibfield  {title} {\bibinfo {title} {{Precision Higgs
  physics at the CEPC}},\ }\href
  {https://doi.org/10.1088/1674-1137/43/4/043002} {\bibfield  {journal}
  {\bibinfo  {journal} {Chin. Phys. C}\ }\textbf {\bibinfo {volume} {43}},\
  \bibinfo {pages} {043002} (\bibinfo {year} {2019})},\ \Eprint
  {https://arxiv.org/abs/1810.09037} {arXiv:1810.09037 [hep-ex]} \BibitemShut
  {NoStop}%
\bibitem [{\citenamefont {Abada}\ \emph {et~al.}(2019)\citenamefont {Abada}
  \emph {et~al.}}]{FCC:2018evy}%
  \BibitemOpen
  \bibfield  {author} {\bibinfo {author} {\bibfnamefont {A.}~\bibnamefont
  {Abada}} \emph {et~al.} (\bibinfo {collaboration} {FCC}),\ }\bibfield
  {title} {\bibinfo {title} {{FCC-ee: The Lepton Collider}: {Future Circular
  Collider Conceptual Design Report Volume 2}},\ }\href
  {https://doi.org/10.1140/epjst/e2019-900045-4} {\bibfield  {journal}
  {\bibinfo  {journal} {Eur. Phys. J. ST}\ }\textbf {\bibinfo {volume} {228}},\
  \bibinfo {pages} {261} (\bibinfo {year} {2019})}\BibitemShut {NoStop}%
\end{thebibliography}%

\end{document}